\documentclass{aastex61}

\def\fp1{\mbox{$a\log(R_e)+b\log(<I_e>)+c\log(\sigma)+d=0$}}
\def\muem{\mbox{$\log(<I_e>)$}}
\def\re{\mbox{$R_e$}}
\def\muere{\mbox{$\log(<I_e>) - \log(R_{\rm e})$}}
\def\muerespace{{$\log(R_e)- \log(<I_e>) - \log(\sigma)$}}
\def\H0{\mbox{$H_0$}}
\def\q0{\mbox{$q_0$}}

\def\eg{{\it e.g.\/}}
\def\ie{{\it i.e.\/}}

\received{December 6, 2016}
\submitjournal{ApJ}

\shorttitle{On the origin of the FP and FJ relations}
\shortauthors{D'Onofrio et al.}

\begin{document}

\title{On the origin of the Fundamental Plane and Faber-Jackson relations: consequences for the star formation problem}

\correspondingauthor{Mauro D'Onofrio}
\email{mauro.donofrio@unipd.it}

\author[0000-0001-6441-9044]{Mauro D'Onofrio}
\affil{Department of Physics and Astronomy, University of Padova, Vicolo Osservatorio 3, I35122 Padova, Italy}

\author{Stefano Cariddi}
\affil{Department of Physics and Astronomy, University of Padova, Vicolo Osservatorio 3, I35122 Padova, Italy}

\author{Cesare Chiosi}
\affil{Department of Physics and Astronomy, University of Padova, Vicolo Osservatorio 3, I35122 Padova, Italy}

\author{Emanuela Chiosi}
\affil{Department of Physics and Astronomy, University of Padova, Vicolo Osservatorio 3, I35122 Padova, Italy}

\author{Paola Marziani}
\affil{INAF - Padova Observatory, Vicolo Osservatorio 5, I35122 Padova, Italy}



\begin{abstract}

The aim of this work is to show that the origin of the Fundamental Plane (FP) relation for early-type galaxies (ETGs) can be traced back to the existence of a fine-tuning between the average star formation rate $<SFR>$ of galaxies and their structural and dynamical characteristics. To get such result it is necessary to imagine the existence of
two distinct "virtual planes" for each galaxy in the \muerespace\ space. The first one (named Virial Plane VP) represents the total galaxy mass using the scalar Virial Theorem and the mass-to-light ratio $M/L$, while the second plane comes from an
expression of the total galaxy luminosity as a function of the mean star formation rate $<SFR>$ and
the velocity dispersion $\sigma$, through a
relation $L=L'_0 \sigma^{-2}$ (named here pseudo-Faber-Jackson (PFJ)) which is a mathematical convenient way for expressing the independency of light from the virial equilibrium. Its validity can be connected to
the mutual correlation $L\sim\sigma\sqrt{<SFR>}$ observed for all ETGs.

A posteriori it is possible to see that this approach permits to explain the observed properties of the FP (tilt and scatter) and the Zone of Exclusions (ZOE) visible in the FP projections.
Furthermore, the link between the properties of the FP and the SFR of galaxies provides a new idea of the star formation, as a phenomenon driven by the initial conditions of proto-galaxies and regulated across the whole cosmic history by the variation of the main galaxy parameters (mass, luminosity, structural shape and velocity dispersion).

\end{abstract}




\section{Introduction}\label{Intro}

The origin of the Fundamental Plane, \ie\ the relation:
\begin{equation}
\fp1
\label{eq1fp}
\end{equation}
between the effective surface brightness, the effective radius and the
central velocity dispersion of early-type galaxies (ETGs), is still unclear
since the epoch of its discovery
\citep{DjorgDavis,Dressetal}. The problem consists in the
observation that the FP coefficients deviate
significantly from the virial expectation for homologous galaxies and in the fact that
the scatter around the plane is very small along the whole FP extension.

The first interpretation of the tilt was related to the behaviour of the stellar populations of galaxies through
their stellar mass-to-light ratio which was seen to vary with luminosity ($M/L \sim M^{\alpha}$, with
$\alpha\sim0.25$ \cite{Faber87}). Subsequent, independent measurements
found similar values of $\alpha$ \citep[see \eg\
][]{Pahre,Gerhard,Borriello,Treu05}.

An alternative explanation was that galaxies are
progressively non homologous systems along the FP
\citep{Hjorth,PrugSimien,Busarello,GrahamColless,Pahre,Bertin,Trujillo,Nipoti,LaBarbera}.
This scenario was supported by the observation that the light profiles and dynamics of ETGs
deviate systematically from homology
\citep{Capaccioli87,deCarvalho88,Capaccioli89,Burkert93,Michard85,Schombert86,Caon,YoungCurrie,PrugSimien}.
\cite{Ciotti} however pointed out that a strong fine--tuning between stellar
mass-to-light ratio ($M^*/L$) and structure (Sersic index $n$) is
required to explain with just structural non-homology both the tilt of
the FP and the small scatter around it (the so-called $M^*/L - n$
conspiracy). \cite{Cappellari,Cappellari3} also
excluded an important contribution of non-homology to the tilt using
integral models of the ETGs mass distribution based on 2D kinematic
maps. Along the same vein, the galaxy mass distribution estimated
from gravitational lensing by \cite{Bolton} did not seem to support an important role for non-homology.

Subsequent interpretations of the tilt proposed a number of possible mechanisms: metallicity effects
\citep{Gerhard}, dark matter distribution and amount (DM) \citep{Tortora,Secco2001,Secco}, dissipation effects during galaxy collapse
\citep[see \eg\ ][]{Onorbe,DekelCox,Robertson,Hopkins}, variable
initial mass function (IMF) \citep{Chiosi}, star formation history (SFH), etc., but
the contribution of DM and IMF was also excluded
by \cite{Ciotti} on the basis of a required strong fine-tuning argument,
and observing that the observed SFH of galaxies is hardly
reconciled with the widely accepted hierarchical paradigm of the
$\Lambda$CDM cosmology.

More recently \cite{Donofrioetal} proposed the existence of a fine-tuning mechanism able 
to explain the properties of 
the FP based on the observed mutual correlation between galaxy mass, mass-to-light ratio 
and Sersic index.

In addition to the tilt the small
observed scatter ($\sim 20-25\%$) around the FP is also unexplained.
\cite{Forbes} and
\cite{Terlevich} found a correlation between the residuals of the
FP and the age of the galaxies (ETGs with higher/lower surface
brightness have younger/older ages).  \cite{Gargiulo} claimed that the
FP residuals anti-correlate with the mean stellar age, while a strong
correlation exists with $[\alpha/Fe]$.  \cite{Gravesetal} proposed
that the stellar population variations contribute at most 50\% of the
total thickness and that correlated variations in the IMF or in the
central DM fraction make up the rest.  \cite{Magoulas} found
that the residuals about the FP show significant trends with
environment, morphology and stellar population, with the strongest
trend being with age.

The above discussion clearly reveals that a general consensus about
the origin of the FP and its properties is still lacking.
We remember that even the distribution of galaxies in the
\muere\ plane, \ie\ one of the projections of the FP, is poorly understood.
\cite{Kormendy} showed that ETGs do not follow the distribution expected for galaxies of 
the same total luminosity,
but are tilted with respect to this line, while \cite{BBF} and \cite{BBFN} noted that 
in this plane
galaxies seem to avoid a region of space: the so called Zone of Exclusion (ZOE).
They claimed that the slope of the ZOE and the progressive displacement of the Hubble 
types from this line is
consistent with the hierarchical clustering scenario with a $n=—1.8$ power-law density 
fluctuation spectrum (plus dissipation).

The same considerations can be done for the Faber-Jackson relation connecting galaxy 
luminosity with
velocity dispersion \citep[$L\propto\sigma^{\sim4}$;][]{FaberJackson},
whose slope (and zero point) changed progressively (today the measured slope is $\sim 2.0$).
This relation is considered a projection of the FP and as such was also related to the 
Virial Theorem, but alternative explanations are possible.

In this paper we propose a new possible solution for the origin of the FP and FJ 
relations able to explain all their observational properties.
The paper is organized as follows: in the first section we present the
main equations and assumptions that define the FP problem.  In
Sec.~\ref{sec2} we describe our proposed solution and in
Sec.~\ref{sec3} we provide the observational evidences in favor of our
hypothesis. In Sec.~\ref{sec4} we discuss the origin of the FJ and PFJ relations
and in Sec.~\ref{sec5} the consequences of our solution for the problem of the star 
formation activity
in galaxies across the cosmic history. Finally in Sec.~\ref{sec6} we
draw our conclusions.

\section{The FP problem}\label{sec1}

We assume that ETGs are gravitationally bound stellar systems which
satisfy the Virial Theorem equation:
\begin{center}
\begin{equation}\label{eqvir}
\langle V^2 \rangle = \frac{GM_{tot}}{\langle R \rangle}.
\end{equation}
\end{center}
where $M_{tot}$ is the total galaxy mass, $\langle R \rangle$ a
suitable mean radius, and $\langle V^2 \rangle$ a mean kinetic energy
per unit mass. By definition every kind of virialized system must
belong to the Virial Plane (VP) in the space defined by the variables
$M_{tot}$, $\langle R \rangle$ and $\langle V^2 \rangle$.
Unfortunately, these are not observable quantities. Therefore, in the
case of ETGs, the Virial Eq. (\ref{eqvir}) is usually written as
follows:
\begin{center}
\begin{equation}\label{eqMtot}
M_{tot} = \frac{K_V\sigma^2\re}{G}
\end{equation}
\end{center}
where $K_V=1/(k_vk_r)$ takes into account projection effects, density
distribution and stellar orbits distribution.  The term $K_V$
parameterizes our ignorance about the orientation, 3D structure and
dynamics of ETGs.  The formal expression of $K_V$ (which is a dimensionless quantity) assumes:
$\langle V^2 \rangle = k_v \sigma^2$, and $\langle R \rangle =
k_r\re$.

Introducing the mean effective surface brightness $\langle I \rangle_e
= L/2\pi \re^2$, one gets such expression for the Virial Plane (VP):
\begin{center}
\begin{equation}\label{eqvir5}
\re = \frac{K_V}{2\pi G}\ (\frac{M_{tot}}{L})^{-1}\ <I_e>^{-1}\ \sigma^2,
\end{equation}
\end{center}
or, in logarithmic form:

\begin{eqnarray}\label{eqvirlog}
\log(\re)&=&2\log(\sigma)-\log(<I_e>)+\log(K_V)+\\ \nonumber
         & &-\log(\frac{M_{tot}}{L})-\log(2\pi G),
\end{eqnarray}

This formulation of the Virial Theorem is directly comparable with the
FP of Eq. (\ref{eq1fp}) rewritten with $\log(\re)$ as independent variable as empirically 
derived from observations.

Note that for a given mass $M_{tot}$ and zero point there are infinite values of \muerespace\
which satisfy Eq. (\ref{eqvirlog}): all the points belonging to a plane obey such equation.
We can therefore define the VP as {\it the locus of points
of the \muerespace\ space which reproduce a constant mass $M_{tot}$ for an assigned zero point.}
In other words the Virial Theorem does not provide any constraints on the position of a galaxy in the \muerespace\  
space.
Two galaxies with the same mass and zero point, but with a different combination of $M/L$ and $K_V$, may share the same VP.
In general Eq. (\ref{eqvirlog}) defines a family of planes filling the \muerespace\ space for all galaxies.

The zero point of Eq. (\ref{eqvirlog}) is given by the quantity:
\begin{equation}\label{eqKV}
ZP_{FP}=\log(K_V)-\log(\frac{M_{tot}}{L})-\log(2\pi G),
\end{equation}
so that each galaxy has its own zero point characterized by a peculiar $M/L$ (dark matter and stellar content) and $K_V$ (degree of non-homology). 
If ETGs were perfectly homologous systems (same $K_V$) with similar $M/L$ the $ZP_{FP}$ would be a constant and all galaxies will be distributed 
along one VP.

In the \muerespace\ space each VP is parallel to the others, so that in principle one should observe a cloud and not a plane, unless some mechanism
constrain all galaxies on the observed FP.

The connection between the FP and the VP clearly links the tilt of the plane to the properties of the stellar population,
to the Dark Matter content and the galaxy structure and dynamics. It is therefore not surprising that all the proposed solutions have tried to
demonstrate the link of the zero point with these galaxy properties. The existence of the FP, with its tilt and small scatter, requires a connection
between $K_V$ (structure) and $M/L$ (DM and stellar populations). This is the so-called fine-tuning problem.

\section{The new proposed solution}\label{sec2}

The new proposed solution comes from the observation that a galaxy of a given mass
$M_{tot}$ has not a defined position in the \muerespace\ space. Its virial equilibrium
is guaranteed by all possible combinations of the variables that fit the virial equation.
It would be nice to have at least another constraint to better
define the location of a galaxy in the \muerespace\ space.

In order to find such constraint we consider that a galaxy of a given mass $M_{tot}$ has
also a total luminosity $L_{tot}$. The luminosity of a galaxy ultimately depends on the luminosities of its stars,
that in turn depend on the star radius and the effective temperature that each star reaches at its surface.

The common way of introducing the luminosity in the FP problem was through the mass-to-light ratio, but we note
that luminosity is actually a quantity independent on the virial equilibrium, being only the product of the SF history of galaxies.

On the basis of such consideration we look for the various expressions that can give the total luminosity of galaxies.
We know that the integrated luminosity $L$ of a galaxy of age $T_G$ can be expressed as:

\begin{equation}\label{Lpop}
L=\int_0^{\infty} \int_o^{T_G} \int_{M_L}^{M_U} S(M,t, Z(t))f_{\lambda}(M,\tau',Z(\tau')) dM dt d\lambda 
\end{equation}

\noindent
where $S(M,t,Z(t))$ is the stellar birth-rate, $f_{\lambda}(M,\tau',Z(\tau'))$ is 
the monochromatic flux of a star of mass $M$, metallicity $Z(t)$ and age $\tau'=T_G-t$, and 
$M_L$ and $M_U$ the minimum and maximum star masses that are formed.
The stellar birth rate $S(M,t,Z(t))$  can be expressed as the total mass converted into stars 
per unit time (e.g. $M_\odot$ $yr^{-1}$) or 
the total number of stars formed per unit time at the time $t$ with the chemical composition 
$Z(t)$. We adopt the first definition for the sake of
consistency with the definition of other quantities in usage here that are related to the star 
formation.
Separating the $S(M,t, Z(t))$ into the product of the SFR $\Psi(t, Z(t))$ and the initial 
mass function $\Phi(M, Z(t))$, and neglecting here
the dependence on the metallicity (it can be easily introduced whenever necessary) 
the above integral becomes

\begin{equation}\label{Lpop1}
L=\int_0^{\infty} \int_o^{T_G}  \Psi(t)F_{\lambda}(\tau') dt d\lambda 
\end{equation}
where
\begin{equation}\label{Lpop2}
F_{\lambda}(\tau')= \int_{M_L}^{M_U} \Phi(M) f_{\lambda}(M,\tau') dM
\end{equation}

\noindent
where $F_{\lambda}(M,\tau')$  is the integrated monochromatic flux at each epoch 
provided by a single stellar population 
of age $\tau'$ and $f_{\lambda}(M, \tau')$ is the monochromatic flux emitted by 
a star of mass $M$ and age $\tau'$ or $t$ in general. Finally, we
define the luminosity per unit mass of a single stellar population as
\begin{equation}
 L_{sp} (t) = \int_0^{\infty} F_{\lambda}(t) d\lambda
\end{equation}
and finally 
\begin{equation}\label{Lpop1a}
L = \int_0^{T_G} \Psi(dt) L_{sp} (t) dt.
\end{equation}
\noindent
We can rewrite Eq. (\ref{Lpop1a}) considering the average values of the involved variables 

\begin{equation}\label{SFR1}
L \,\, \sim \,\, <\Psi(t)\times L_{sp} > T_G     
\end{equation}

\noindent
where $<\Psi(t) \times L_{sp} >$ is the time averaged product of the current 
SFR and the luminosity of  the  stellar populations, 
$T_G$ is the  age of the galaxy. In the above average, $L_{sp}$  indicates   the mean stellar 
population representative 
of the whole stellar content. The emitted light is per unit mass.  
Eq. (\ref{SFR1}) is substantially telling us that the total luminosity of galaxies is the result of its SFH.

We also know however that the luminosity of ETGs is observed to correlate with the velocity dispersion
of their stars through the Faber-Jackson relation \citep{FaberJackson}.

\begin{equation}\label{LFJ}
L=L_0\sigma^{\beta}
\end{equation}
with $\beta \sim 2-4$. The origin of this correlation is obscure.

In the following we leave the expression in this form instead of scaling it in the form
$L=L_0(\sigma/\sigma_0)^{\beta}$ because we want to emphasize the physical meaning of
the parameter $L_0$ whose units are $[gr/sec]$ consistent with a SFR if $\beta=2$. 

\begin{figure}[ht!]
\plotone{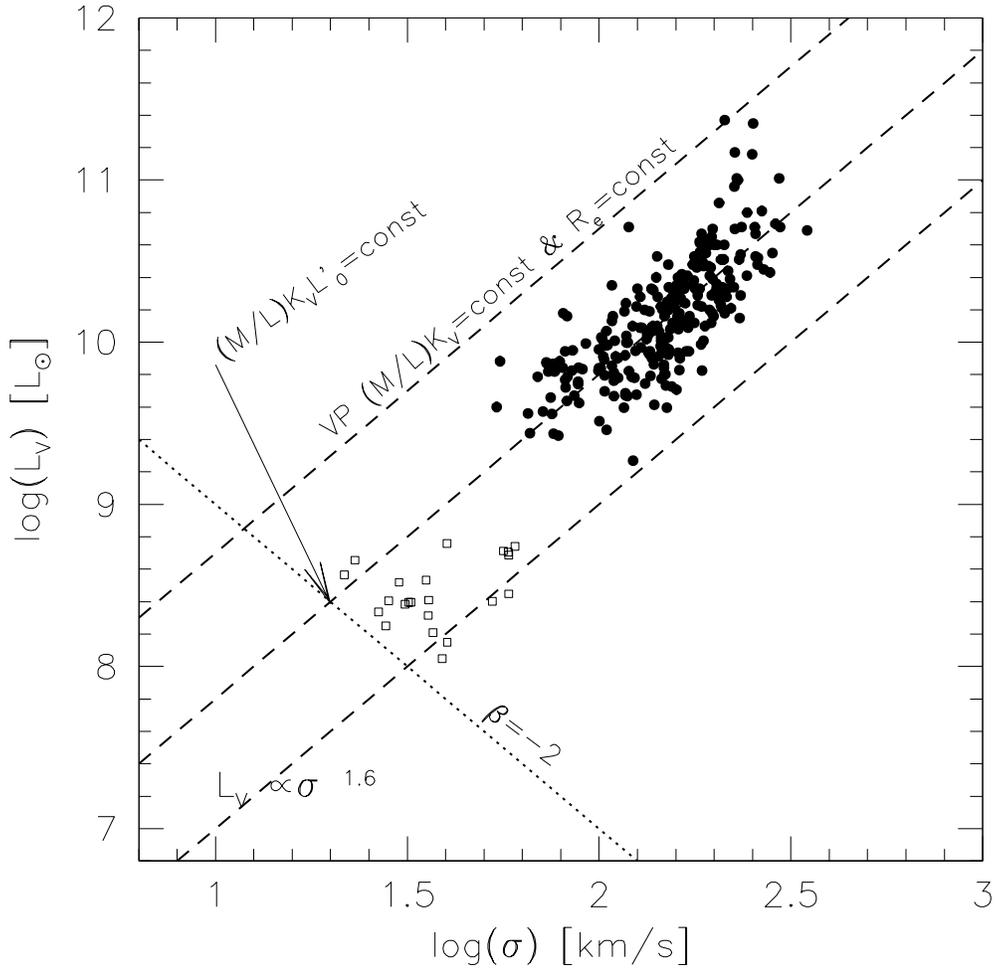}
\caption{The $L-\sigma$ plane. The fitted FJ relation for the ETGs of the WINGS database \citep{Morettietal} is given by the solid line.
The dashed lines mark the position of the VPs for galaxies with different effective radii and zero-point. The dotted line marks 
one possible PFJ plane with $L'_0=constant$
with slope equal to $-2$ (see text). The classical FJ relation seems to result from the intersection of the PFJ and the VP planes.
The filled circles are normal ETGs. The open squares are dwarf galaxies of the WINGS database with masses around $10^8-10^9 M_{\odot}$.\label{Fig5}}
\end{figure}

The fundamental question is why the galaxy dynamics should be aware of the stars that have been produced across the cosmic time.

Up to now the FJ relation has been considered a direct consequence of the Virial relation for systems where the mass-to-light ratio 
$M/L$ vary systematically with the galaxy mass or luminosity. We will see below that alternative explanations are possible.

The direct comparison of Eq. (\ref{SFR1}) and Eq. (\ref{LFJ})
tells us that the parameter $L_0$ of the FJ relation
is connected to the mean SFR. We can in fact write:

\begin{equation}\label{SFR2}
L_0=<\Psi(t) \times L_{sp}> T_G/\sigma^2.
\end{equation}

In this parameter is encrypted the complex
relationship between the galaxy dynamics and the SFH.

\section{The observed projections of the Fundamental Plane}\label{sec3}

What can we say observationally? Could we demonstrate the existence of a link between the Virial and FJ planes giving rise to
the FP tilt? We will see here that this is not the case if $\beta=2$.

From the observational point of view it is better to look at the projections of the FP, \ie\ at the \muere\ plane, the $\muem-\log(\sigma)$ plane and the
$\log(\sigma)-\log(R_e)$ plane.

The question is: where are located the projections of the intersecting lines, \ie\ the lines of constant $M/L$, $K_V$ and $L'_0$ in these 2D planes?

In order to answer such question we should consider Eqs. (\ref{eq1fp}), (\ref{eqvirlog}) and 
to remember that $L_{tot}=2\pi<I_e>R^2_e$, so that
passing to the logarithms Eq. \ref{LFJ} can be rewritten:

\begin{eqnarray}\label{eqBB4}
\log(R_e) &=& (\beta/2)\log(\sigma) - (1/2)\log(<I_e>) +\\
          & &+(1/2)\log(L_0)-(1/2)\log(2\pi).\nonumber
\end{eqnarray}

The same equations can be also written as a function of $\sigma$ in the following way:
\begin{eqnarray}\label{eqsys}
\log(\sigma)&=&A\log(\re)+B\log(<I_e>)+C\nonumber  \\
\log(\sigma)&=&\frac{1}{2}\log(\re)+\frac{1}{2}\log(<I_e>)+\frac{1}{2}\log(M/L)+\nonumber \\
            & &-\frac{1}{2}\log(K_V)+\frac{1}{2}\log(2\pi G) \\
\log(\sigma)&=&\frac{2}{\beta}\log(\re)+\frac{1}{\beta}\log(<I_e>)+\nonumber \\
            & &-\frac{1}{\beta}\log(L_0)+\frac{1}{\beta}\log(2\pi)\nonumber
\end{eqnarray}
where the coefficients $A$, $B$ and $C$ are related to those of Eq. (\ref{eq1fp}).
Then we take the difference FP-VP and FP-FJ. These differences must be equal on the intersecting lines.
It follows after some algebra that:

\begin{equation}\label{eqsys1}
\log(<I_e>)=\frac{(2/\beta)-(1/2)}{(1/2)-(1/\beta)}\log(\re)+  \Pi
\end{equation}
where $\Pi$ contains all terms not explicitly written in the Eq. (\ref{eqsys1}).

Now we ask ourself if
Eq. (\ref{LFJ}) could represent the plane we are looking for
in the \muerespace. First we observe that in the FJ relation $L_0$ is nearly constant for almost all ETGs (in the mass range $10^9 - 10^{12} M_{\odot}$)
of different $\sigma$. So this relation is not the one we are looking for as a second virtual plane
representing the total luminosity of a galaxy in the \muerespace\ space.
Furthermore for $\beta=2$ Eq. (\ref{eqsys1}) the slope of the $I_e-R_e$ relation is undefined.

Looking at Fig. \ref{Fig5} we
note instead that an alternative way of writing $L_{tot}$ is possible
and mathematically correct:

\begin{equation}\label{SFR3}
L=L'_0\sigma^{\beta}
\end{equation}
where the value of $\beta$ could be chosen on the basis of the observed distribution of galaxies
in the FP projections. We will see that the best value for $\beta$ is $-2$. The slope of such relation is
marked by the dotted line in Fig. \ref{Fig5}.

With such a relation we assign to $L'_0$, which is very different from galaxy to galaxy, 
the primary role of capturing the SFH of each object leaving to $\sigma$ the secondary 
role of indicating how the velocity dispersion affects the SFR
($\sigma$ could only change in a limited interval, that provided by the scatter of the FJ 
relation).

Being $L'_0$ and $L_0$ correlated we have that $L_0=L'_0\sigma^{-4}$.
It follows on the basis of Eq.(\ref{SFR1}) that also $L'_0$ is connected to the SFR:

\begin{equation}\label{SFR4}
L'_0=<\Psi(t) \, L_{sp}> T_G\sigma^2.
\end{equation}

Now substituting $L'_0$ to $L_0$ in Eq. \ref{eqBB4} we obtain 
a plane in the \muerespace\ space which is tilted in the right direction with respect 
to the VP and with the notable property of having a significantly different zero-point for each galaxy.

This is the second virtual plane of the \muerespace\ that we are looking for. It represents 
the total luminosity of a galaxy with a zero-point different for each object as it is the case for the total mass in the VP (through $(M/L)$ and $K_V$ as zero-points).

We call this plane the "PFJ plane" (pseudo-FJ) for keeping in mind its origin from the FJ relation and we define it as follows: 
{\it The PFJ plane is the locus of points
defined by the values of \muerespace, which reproduce a constant luminosity $L_{tot}$ for an assigned zero point $L'_0$.}
This plane contains, as the VP, only one galaxy and all PFJ planes are parallel each other in the \muerespace\ space.

The different inclination of the VP and PFJ planes suggests that they intersect somewhere in the \muerespace\ space, forming a line 
in such space. Along this line it resides only one object, that with mass $M_{tot}$, luminosity $L_{tot}$ and
zero points $Z_{FP}$ and $Z_{PFJ}=1/2\log(L'_0)$. In other words along this line, the product $(M/L)K_VL'_0$ is constant.

It is clear that if the zero points of the VP and PFJ planes vary in a coordinated way, the result will
be that of forming several parallel lines in the \muerespace\ space, each one containing one galaxy. The plane best fitting
this distribution of parallel lines is the plane of real galaxies in the \muerespace\ space, \ie\ the FP.
We therefore define the FP as follows: {\it The FP is the plane
in the \muerespace\ space that best fit all the parallel lines formed by the intersections of the VP and PFJ planes. In this plane the
quantity $(M/L)K_VL'_0$ is constant.} In this framework,  the existence of a FP for real galaxies implies that
a close connection must exist between $(M/L)$, $K_V$ and $L'_0$
(or in other words between mass, luminosity, structure and SFR).

\begin{figure}[ht!]
\gridline{\fig{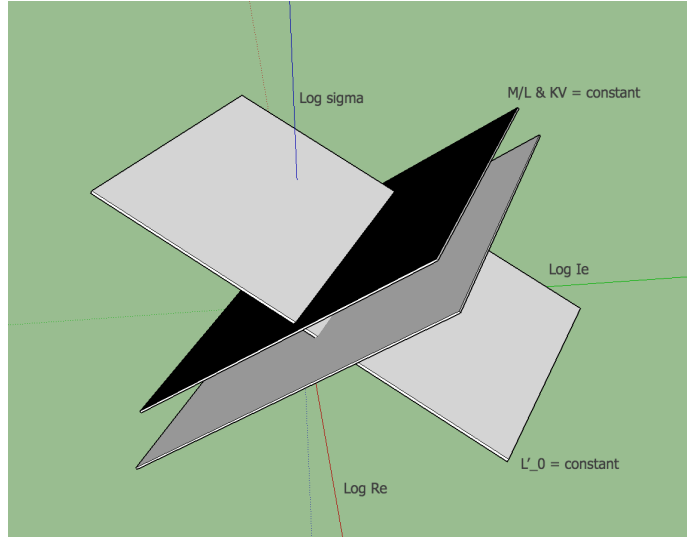}{0.5\textwidth}{(a)}
         }
\gridline{\fig{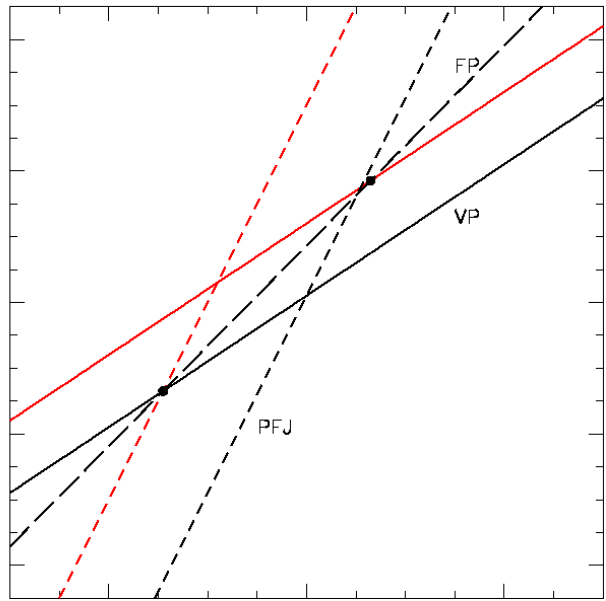}{0.6\textwidth}{(b)}
          }
\caption{Panel (a): General view of the \muerespace\ with two
VPs and one FPJ plane.
Panel (b):Two possible VP and FPJ planes seen edge-on for two ETGs of masses $M_1$ and $M_2$ and luminosities $L_1$ and $L_2$ respectively
are shown with black (VP) and red lines (FPJ).
The FP results in this case from the connection of the two intersections of the VP and PFJ planes. For many galaxies
the FP is the plane best fitting all the intersecting lines.\label{Fig1}}
\end{figure}

A graphical representation of the mechanism originating the FP is given in Fig.\ref{Fig1}. The upper panel of the figure shows two VPs for
two galaxies (in black and gray) and one PFJ plane for one galaxy. The intersecting line formed in the \muerespace\
by the two planes for a galaxy of mass $M_{tot}$ and luminosity $L_{tot}$ marks the locus in which galaxy might reside.

Consequently, the FP plane is naturally tilted with respect to both the VP and PFJ planes. Its tilt is now connected to the global variation of the
zero points of the VP and PFJ planes ($Z_{FP}$ and $Z_{PFJ}$), and
the small scatter observed around the plane originate from the fine-tuning effect linking
$M/L$, $K_V$ and $L'_0$, i.e. linking the galaxy mass, structure and dynamics with the SFR of galaxies.
In view of the future use it is mathematically convenient to assume:

\begin{eqnarray}
\Pi&=&\frac{1}{2}\log(K')=\frac{[\frac{1}{2}\log(K_V)-\frac{1}{2}\log(M/L)}{[\frac{1}{2}-\frac{1}{\beta}]}+\nonumber\\
   & &+\frac{-\frac{1}{\beta}\log(L'_0)-\frac{1}{2}\log(2\pi\ G)+\frac{1}{\beta}\log(2\pi)]}{[\frac{1}{2}-\frac{1}{\beta}]}.
\label{eqzp}
\end{eqnarray}
which also defines the constant $K'$.

We have obtained an equation for the distribution of galaxies with similar $M/L$, $K_V$ and $L'_0$ in the \muere\ relation. The zero point of 
Eq. (\ref{eqsys1})
varies as $M/L$, $K_V$ and $L'_0$ vary in the FP space.
Note that the slope of the relation depends only on the value of $\beta$, \ie\ on the exponent of the PFJ plane.

\begin{figure}
\plotone{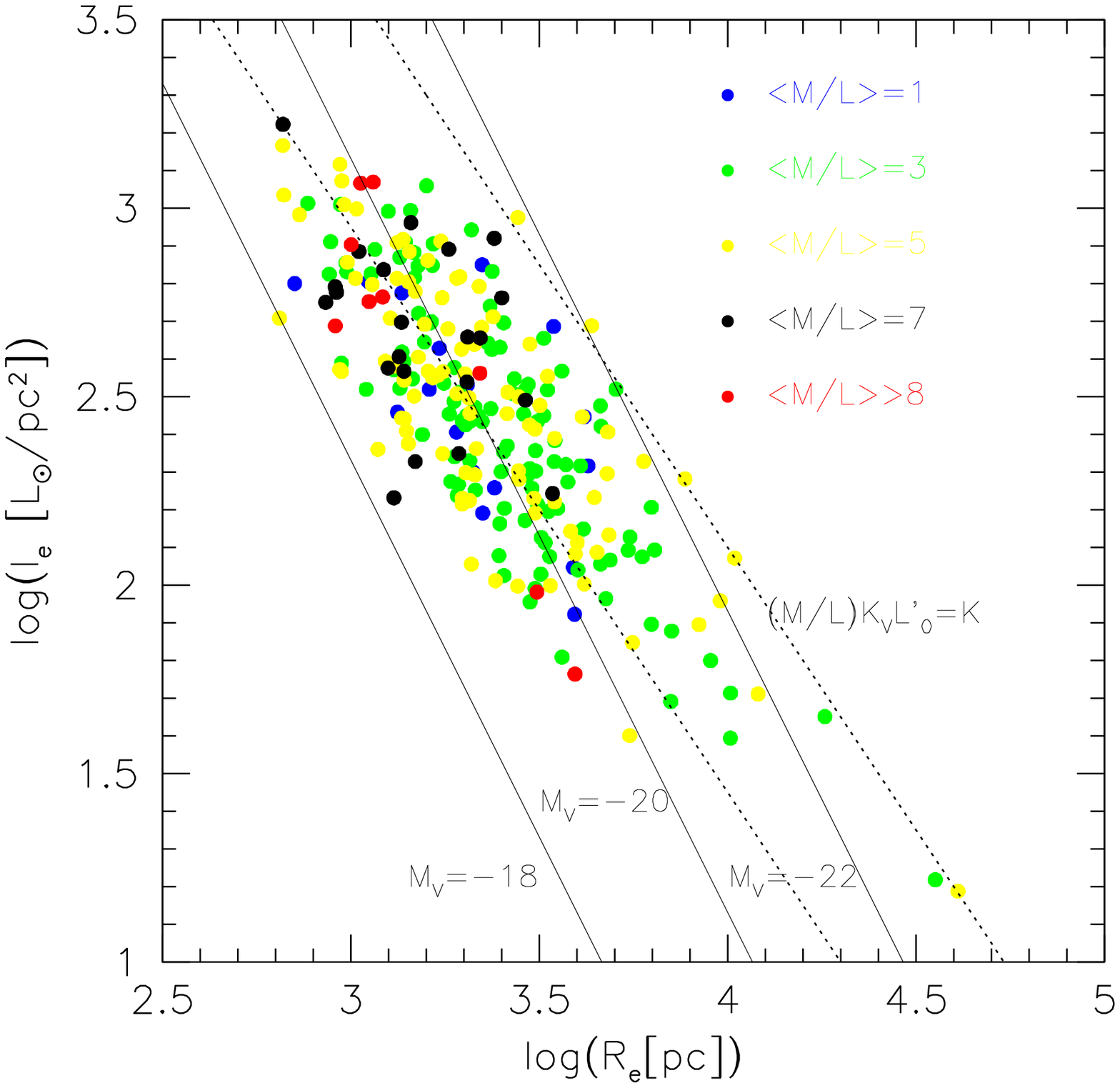}
\caption{The \muere\ plane of the WINGS ETGs. Galaxies are plotted with different colors according to their measured stellar $M^*/L$ as indicated.
The solid lines give the locus of constant galaxy luminosity.
The dotted lines mark the locus of constant $M/L$, $K_V$ and $L'_0$, \ie\ the projections of the intersecting lines originating the FP.\label{Fig2}}
\end{figure}

Fig.\ref{Fig2} shows the \muere\ plane where we have adopted the solution of Eq. (\ref{eqzp}) with $\beta=-2$.
Note how this value of $\beta$ naturally reproduces the slope of the observed distribution of galaxies. It follows that the so called ZOE 
(Zone of Exclusion) is in this context a natural limit reached today by the values of $M/L$, $K_V$ and $L'_0$ during the cosmic evolution.

In the figure we plotted with different colors different ranges for the stellar $M^*/L$ ratios available for the galaxies of the WINGS database 
in the V-band \citep{Morettietal}. Note that
there is not a clear trend in the $M^*/L$ ratios, even if
the higher mass-to-light ratios
seem more frequently distributed far from the ZOE.

\begin{figure}
\plotone{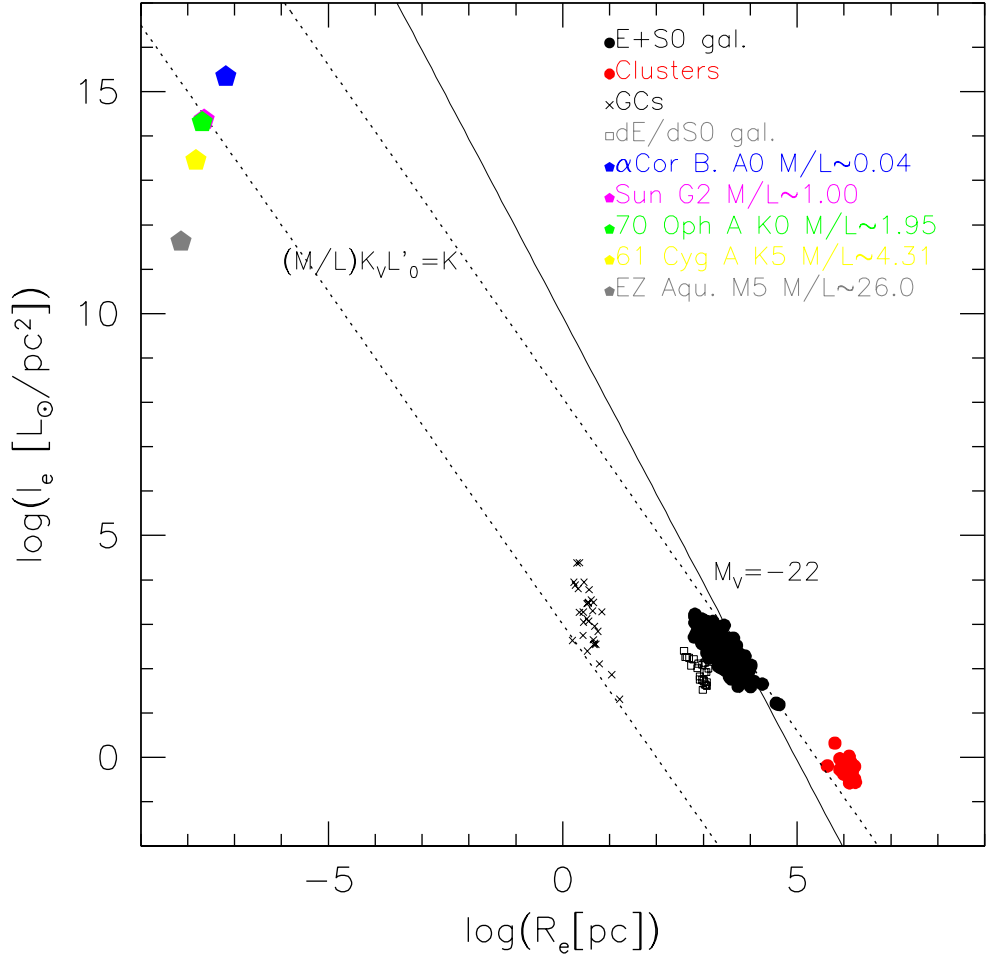}
\caption{The \muere\ plane for objects of different masses that are known to be close to the virial equilibrium: main sequence stars, globular
clusters, dwarf galaxies, normal ETGs and galaxy clusters.
The solid lines give the locus of constant absolute magnitude, while the dotted lines are parallel to the ZOE. The lower dotted line marks the 
position of $M/L=1$ ($M_{\odot}/L_{\odot}$).\label{Fig3}}
\end{figure}

Fig.\ref{Fig3} is instead a plot of the \muere\ distribution for objects of very different masses, covering a range from $\sim1 M_{\odot}$
to $\sim10^{14} M_{\odot}$, i.e. from stars to clusters of galaxies. 
The data for the GC systems are taken from \cite{Pasquato}, those for stars are taken from Wikipedia and that for dwarf galaxies and galaxy 
clusters come from the WINGS database (Cariddi et al. in prep.). 

Note that the \muere\ relation seems to be valid on all scales.
For stars the $M^*/L$ ratio increases as far as we move away from the ZOE going from the main sequence stars of A spectral type to that of M type stars.
If the dominating stellar population inside a stellar system is made of late type stars we will observe
an higher ($M^*/L$) that will likely place the galaxy far from the ZOE\footnote{Assuming that the DM contribution
is approximately the same for all galaxies, which is not exactly the case.}.

Note also that this diagram is done for the V-band, so that there is a natural selection effect working on, since the lower $(M^*/L)$ due to bright 
stars that dominate the galaxy luminosity, progressively move the galaxies toward the ZOE. 

The galaxy clusters appear shifted with respect
to the ZOE because these systems contain several spiral galaxies with low $M^*/L$, while Globular Clusters
have a solar mass-to-light ratio because their stellar population is dominated by stars with high $(M^*/L)$.

\begin{figure}
\plotone{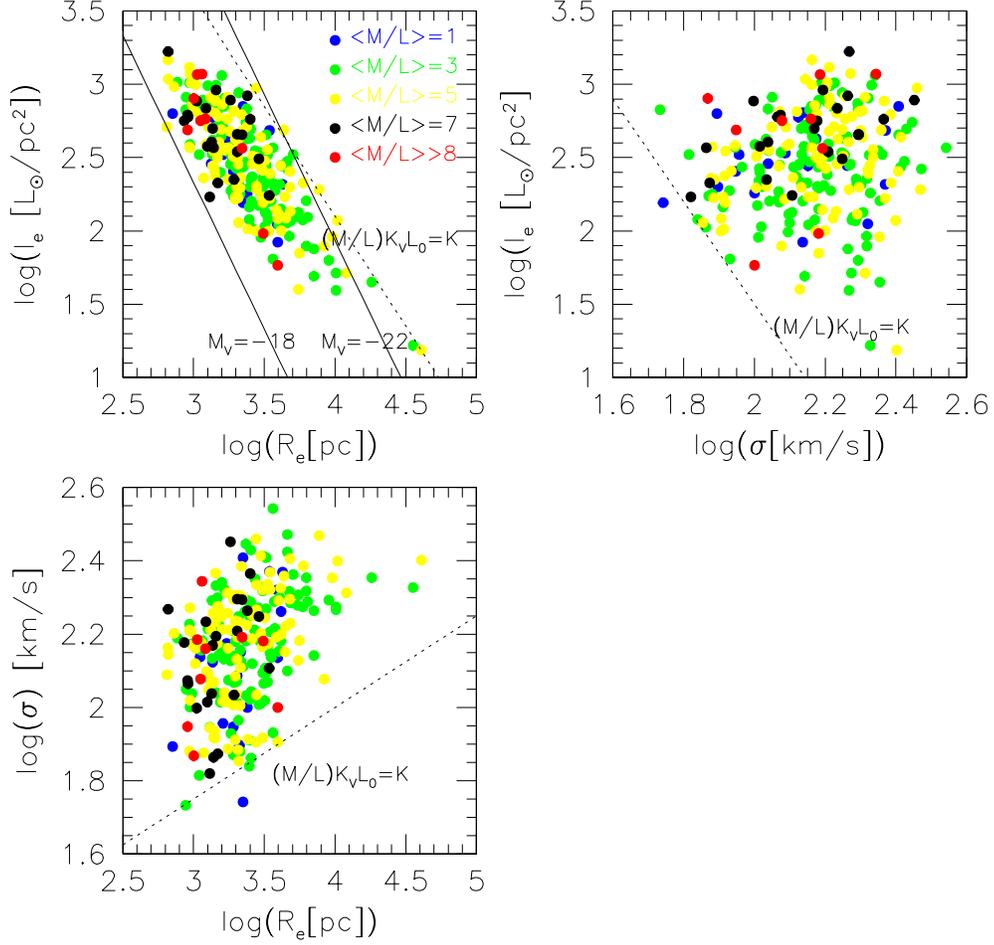}
\caption{The different projections of the FP on the \muerespace\ axes.
The dotted lines mark a possible position for the ZOE.\label{Fig4}}
\end{figure}

For the other FP projections we obtain:
\begin{eqnarray}
\log(<I_e>)&=&(\beta-1)\log(\sigma)+const \nonumber \\
\log(\sigma)&=&\frac{1}{2-\beta}\log(\re)+const,
\label{eqfppro}
\end{eqnarray}
where the constant zero points also depend on the combination of $M/L$, $K_V$ and $L'_0$.
Again the $\beta=-2$ value determines the distribution of galaxies and the position of the ZOE in the respective diagrams (see Fig.\ref{Fig4}).

In conclusion, we have obtained the FP by fitting in the \muerespace\ space the distribution of the parallel lines where for each galaxy two virtual planes intersect each other. The first plane is provided by the Virial Theorem and fixes the mass 
of a galaxy 
once the $M/L$ and $K_V$ zero point are given. The other comes
from having written the total galaxy luminosity with
the $L=L'_0\sigma^{-2}$ relation, encrypting in the parameter
$L'_0$ the role played by the SF activity.

In the next section we will further discuss the possible origin of the connection between luminosity and velocity dispersion in ETGs and the nature of 
the $L_0$ and $L'_0$ parameters.

\section{More on the FJ and PFJ planes}\label{sec4}

Why $L$ and $\sigma$ are correlated variables? A priori
there are no reason at all for such a connection. A posteriori we understand it on the basis of the connection between mass and luminosity in 
each single star and on the basis of the virial theorem. The SF
is a local phenomenon originating by micro-physical processes inside clouds of gas and dust, while the velocity dispersion is a direct consequence 
of the mass potential well. How the two things communicate? This a classic example of a recurrent problem in physics concerning the connection between 
microscopic and macroscopic phenomena.

Before attempting any possible answer we want to better describe here
the $L-\sigma$ plane, which is actually very different from the VP. The FJ plane contains two measured quantities,
the galaxy luminosity and the stellar velocity dispersion. At variance with the VP that is defined for one galaxy only in the \muerespace\ 
assigning its mass
and zero point, the FJ plane contains all real galaxies at the same time. Along the fitted relation the zero point $L_0$  is nearly constant 
for almost all galaxies (let say between $10^9$ and $10^{12}$ $M_{\odot}$).

The first thing to note is that in the FJ plane the points of constant $M/L$, $K_V$ and $L'_0$ are the galaxies themselves (see again Fig.\ref{Fig5}).
Note how the selected solution with $\beta=-2$ used for the \muere\ relation gives here the series of parallel zero points that for each $\sigma$
provides the luminosities of all galaxies
reproducing the observed FJ relation when they are considered all together. The FJ relation seems to originate from the intersections of the "projections" in the $L-\sigma$ space 
(having collapsed $I_e$ and $R_e$ in the variable $L$) of all the parallel virtual planes that represent the total luminosity of galaxies 
with the "projections" arising from the virial planes (the dashed lines where $M/L$, $K_V$ and $R_e$ are constants).
The intersection of
the $L=L'_0\sigma^{-2}$ line with the VP projection fixes the exact position of a galaxy in the $L-\sigma$ space.

The result is the relation expected for all virialized stellar systems having similar zero-point $L_0$. 
In this context it is therefore possible to explain why the residuals from the FJ 
relation correlate
with the $(M/L)$ ratio \citep{Cappellari} and with galaxy sizes \citep{Desmond}.

\begin{figure}
\plottwo{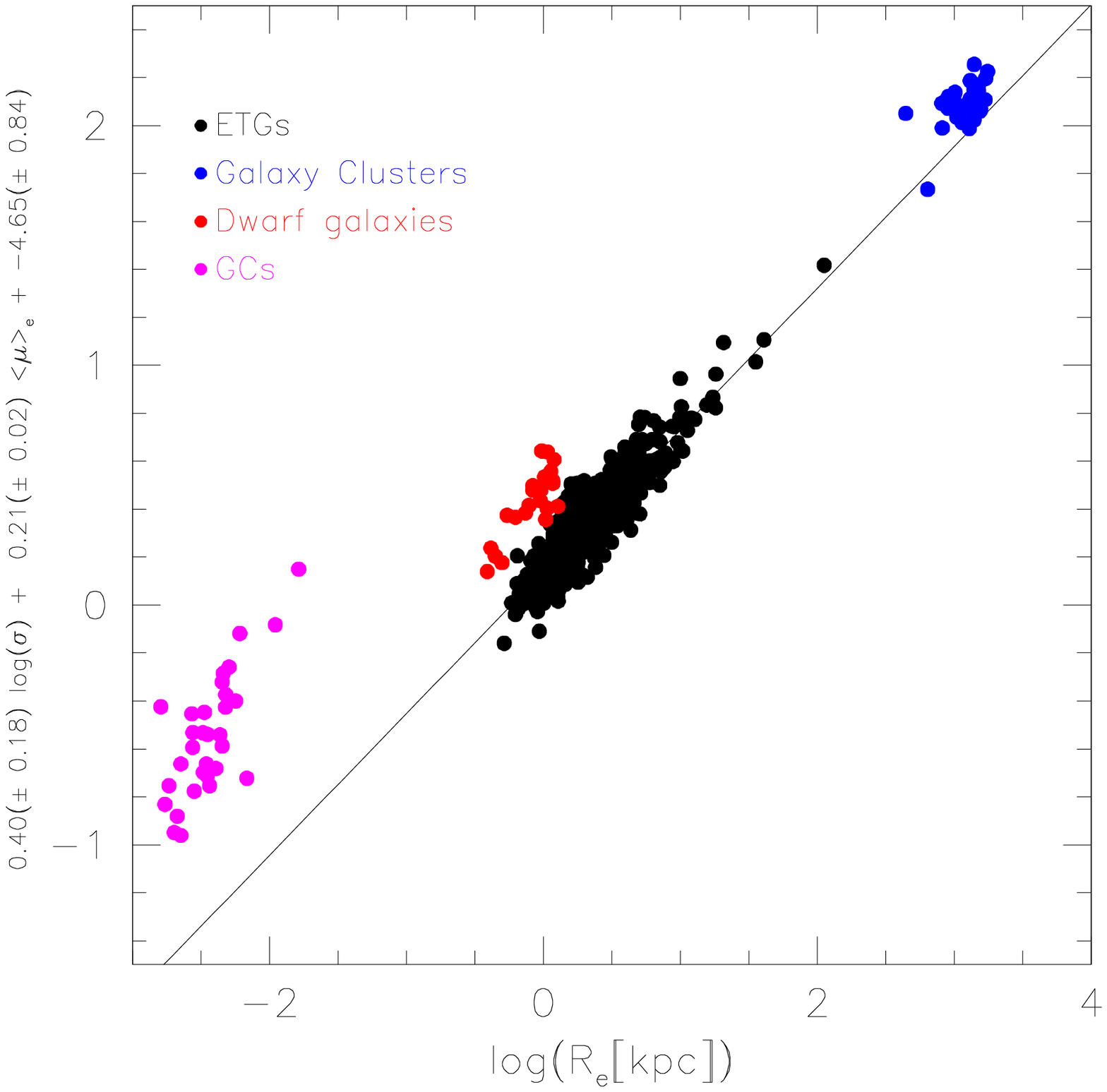}{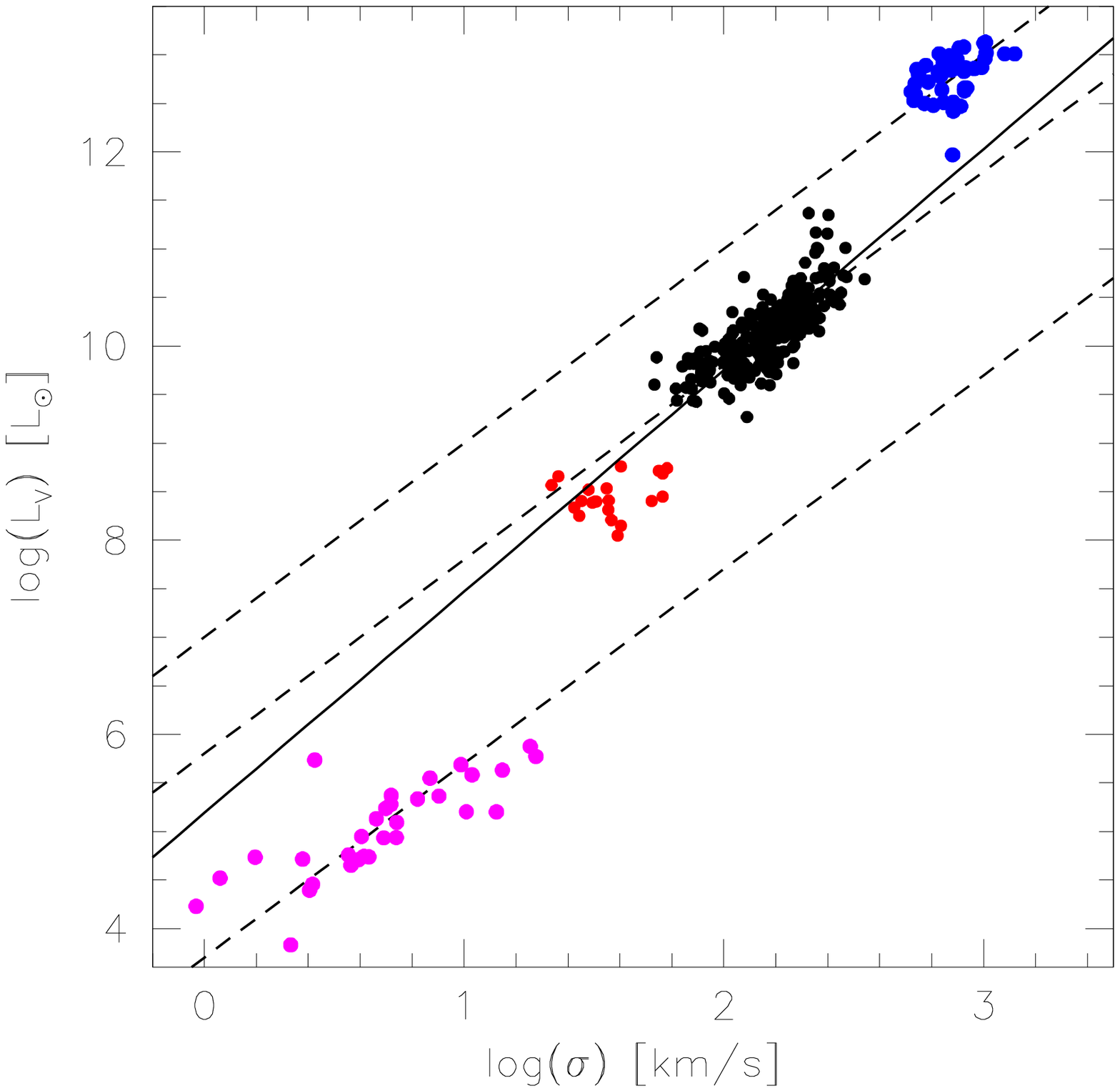}
\caption{Top: an edge-on view of the FP for all types of stellar systems. Bottom: the same systems on the FJ relation.\label{Fig6}}
\end{figure}

Note also that dwarf galaxies ($M\sim10^{8\div9} M_{\odot}$) deviate from the main galaxy relation. This occurs
for the same reason why these stellar systems deviate from the FP (see Fig.\ref{Fig6}): they have a zero point
systematically different, \ie\ different $M/L$, $K_V$ and $L'_0$ values. The upper and lower panels of this figure clearly show that all 
stellar system seem to obey to the FP and FJ relations, but with zero points slightly different from that of typical galaxies. These variations
are responsible of the larger exponent observed in many cases for the FJ relation (4 instead of 2), which ultimately depends on the heterogeneity 
of the galaxy sample, \ie\ from the inclusion of galaxies of very different masses and zero-points. An extreme example is seen in the right panel of Fig.\ref{Fig6}, where a steeper slope for 
the FJ can be obtained with a fit for objects of very different masses (and zero-points).

However, the FJ law is a relation that provides a further element to the virial relation, linking mass (and the virialized system internal gravitational
energy) to the production of radiant energy i.e., to the object luminosity. The mechanisms of production of energy can be very different and can 
yield to widely different $M/L$ even among stellar systems, where the mechanism is roughly the same, ultimately associated with nuclear reactions
in the star interior.

If we now take Eq. (\ref{eqzp}) with $\beta=-2$, giving the zero point of the relationship between the effective surface brightness $I_e$ 
and the effective radius $R_e$ (the zero point varies with $M/L$, $K_V$ and $L'_0$ and hence with $Z_{FP}$),  after few steps we get:

\begin{equation}\label{loki}
K'=\frac{K_V}{4\pi^2 G}\frac{L}{M}L'_0
\end{equation}
where $L'_0$ is $L/\sigma^{-2}$ and
$K'$ is a parameter different for each cosmic epoch with units of $[gr^2 cm^{3} sec^{-6}]$ (or $[L^2_{\odot}/pc]$), the gravitational constant is given in cgs units or expressed as $G=4.3\times 10^{-3} pc M^{-1}_{\odot}$ $(km/s)^2$ and the term $K_V$ is a 
function of the Sersic index $n$ \citep[see,][]{Bertin}.
$K'$ will follow the evolution of the main galaxy parameters by changing the position of a galaxy in the \muere\ plane. As a consequence the whole FP is expected to vary
its tilt across the cosmic epochs.

\begin{figure}
\plotone{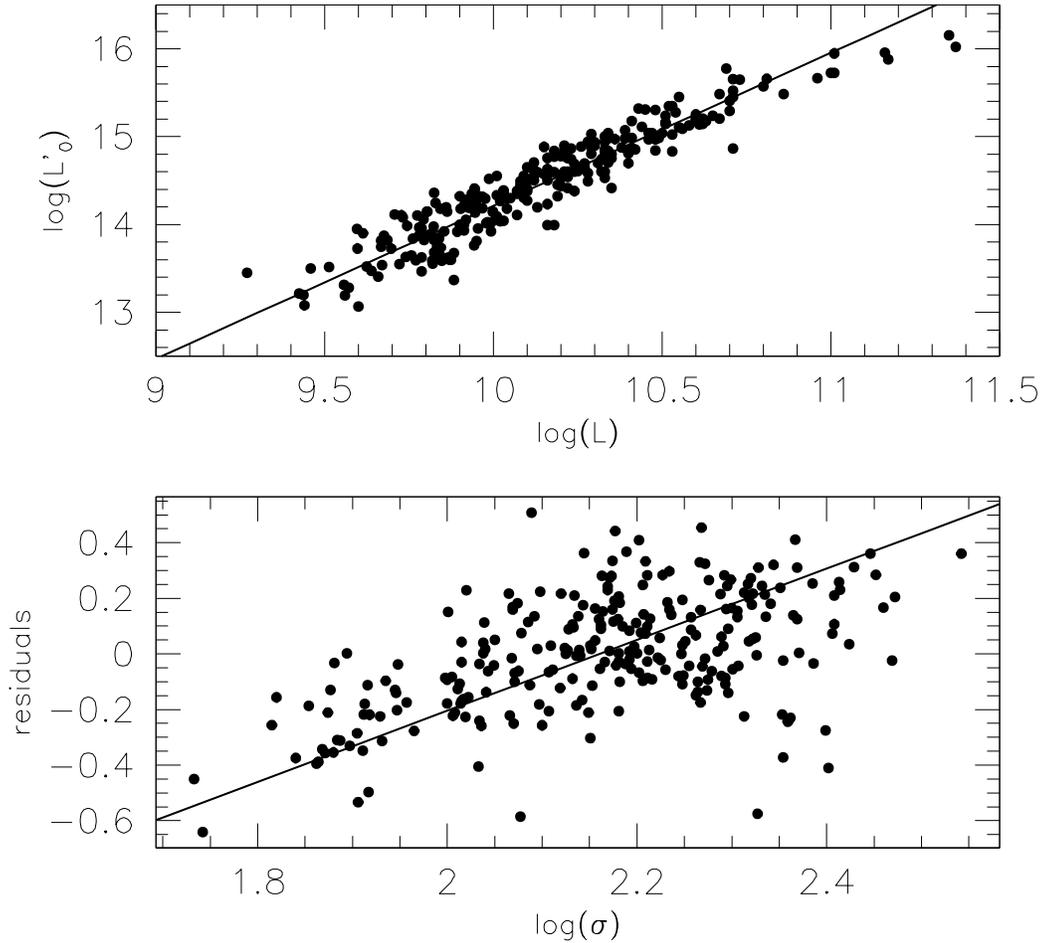}
\caption{Upper panel: Plot of $L'_0$ derived from Eq. \ref{loki} vs the measured total galaxy luminosity $L$. Lower panel: Plot of the residuals from the best fit of the above relation versus the measured velocity dispersion $\sigma$. \label{Fig8}}
\end{figure}

The quantity $L'_0$ is the zero-point of the $L=L'_0\sigma^{-2}$ relation. It 
is marked by the dotted line in Fig. \ref{Fig5} in the FJ space, and intersects the VP lines originating the observed FJ law.
Fig.\ref{Fig8} shows the relation between $L'_0$ derived from Eq. (\ref{loki}) and the total galaxy luminosity $L$. 
Here we used the stellar $M^*/L$ being $M/L$ unknown. We observe that
the link of $L'_0$ and 
$L$ is far from being trivial ($L'_0$ results from a complex combination of $M/L$ and $K_V$), and the same could be said for the dependence of the residuals to the
central velocity dispersion $\sigma$.

\begin{figure}
\plottwo{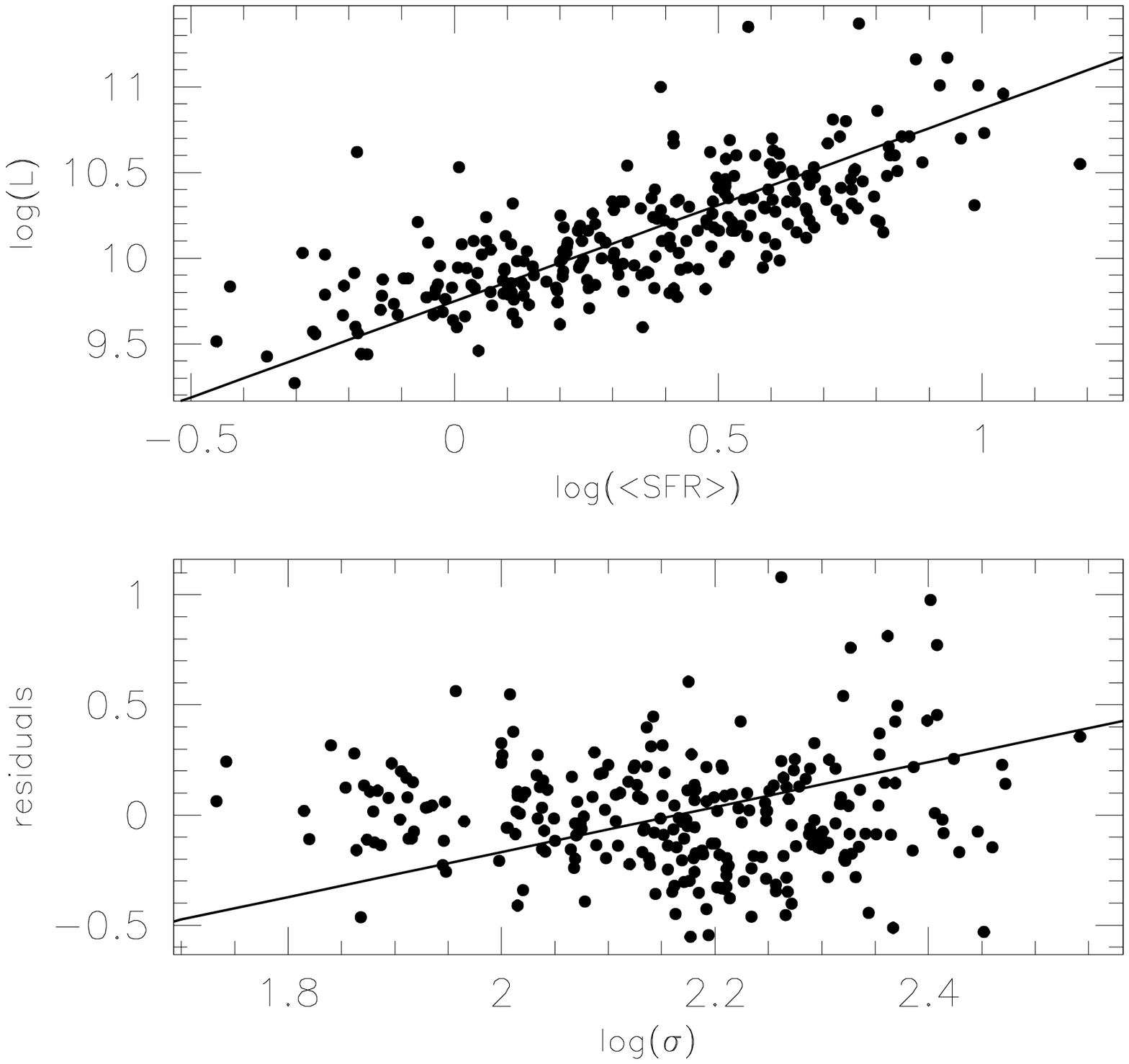}{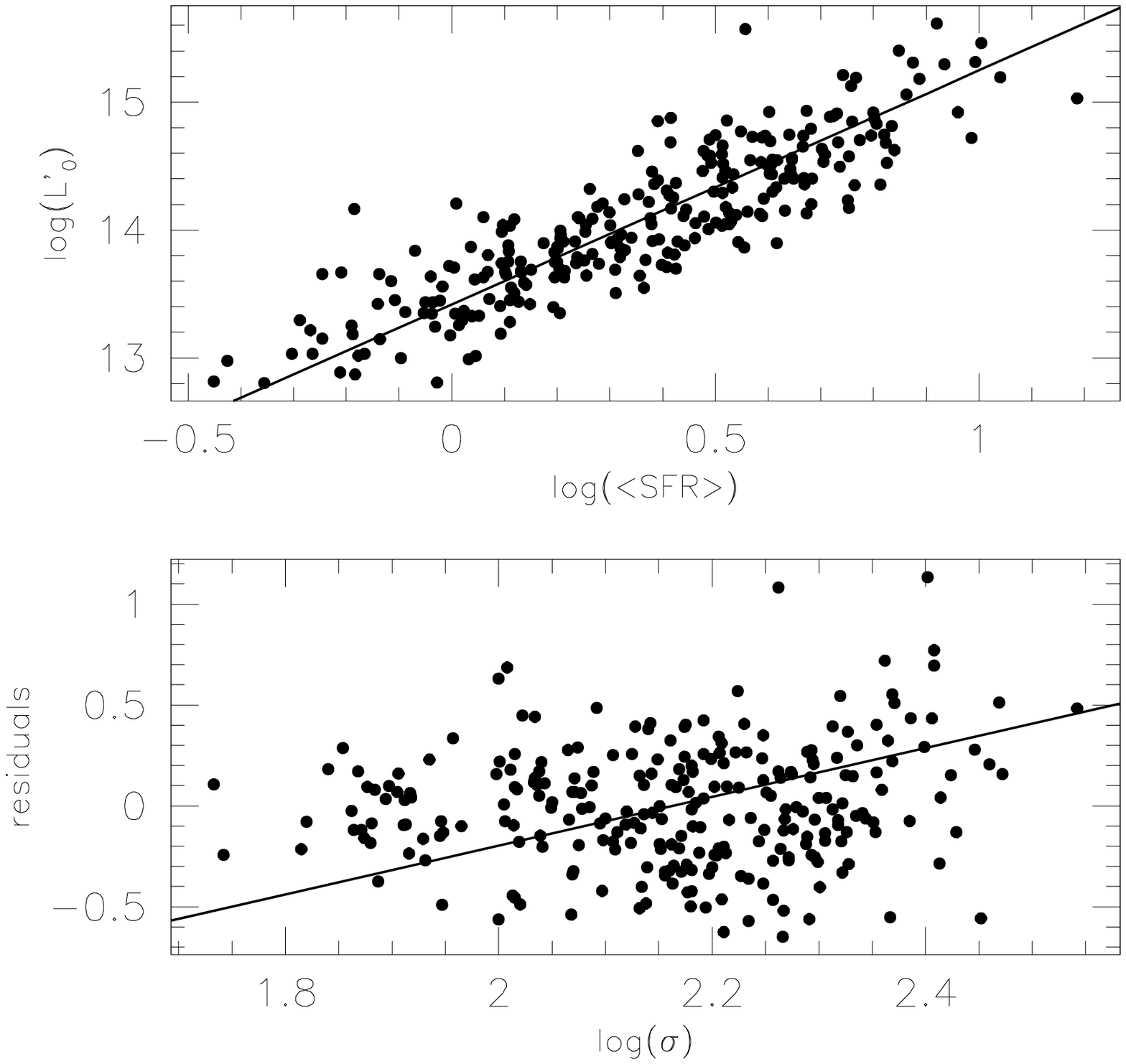}
\caption{Plot of $L$ and $L'_0$ vs the mean SFR in log units. Note that the residuals of these relations depend on $\sigma$.\label{Fig8bis}}
\end{figure}

\begin{figure}
\plottwo{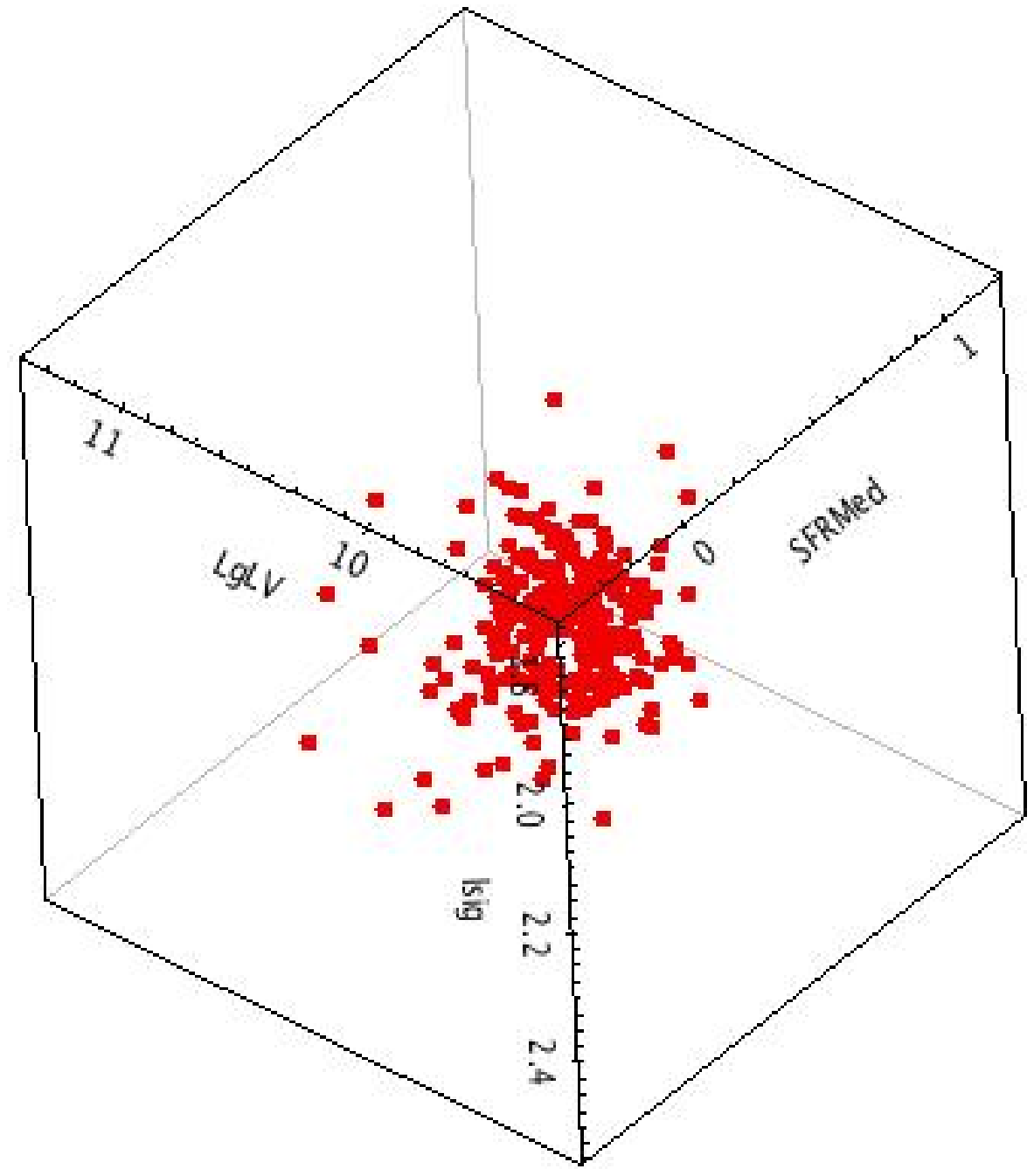}{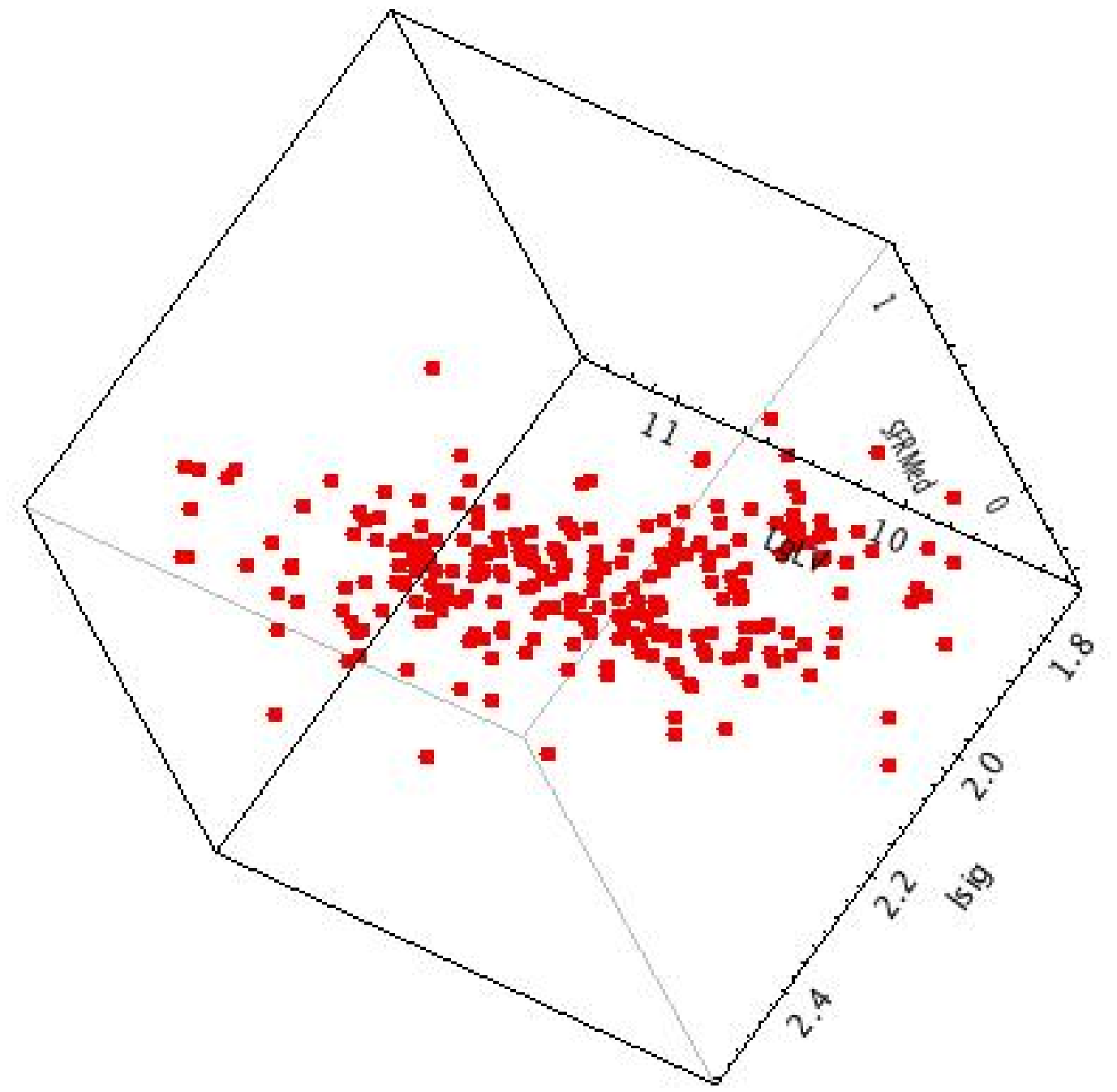}
\caption{Two 3D views of the correlation between $L$, $<SFR>$ and $\sigma$ in log units.\label{Fig8ter}}
\end{figure}

Fig. \ref{Fig8bis} gives a clear indication that both $L$ and $L'_0$ are correlated with the mean SFR of the galaxies measured by \cite{Fritz1}.
The residuals present a significant dependence on $\sigma$. All these things tell us that we should look at the correlation of the three
variables $L$, $\sigma$ and $<SFR>$. These are mutually connected because the mass $M$ correlate with the velocity dispersion $\sigma$
through the virial relation and the light $L$ correlate with the mean star formation rate $<SFR>$. Consequently $\sigma$ and $<SFR>$ are connected.
Fig. \ref{Fig8ter} provides two angle views of the 3D distribution of such variables. Note the elongated sigar-shape distribution of ETGs
in this space. 

The 3D correlation between these variables gives:
\begin{equation}\label{SFRsL}
\log(L)=0.48(\pm0.06)\log(<SFR>)+1.00(\pm0.13)\log(\sigma)
\end{equation}
with an $rms=0.215$ ($R=0.64$ and $p-value<1.2\times10^{-16}$). Therefore $L\sim\sigma\sqrt{<SFR>}$. It is clear in this context that
the $L=L'_0\sigma^{-2}$ relation represents the most convenient way of assigning a role to the SFR of each galaxy in the $L-\sigma$ plane.

\begin{figure}
\plotone{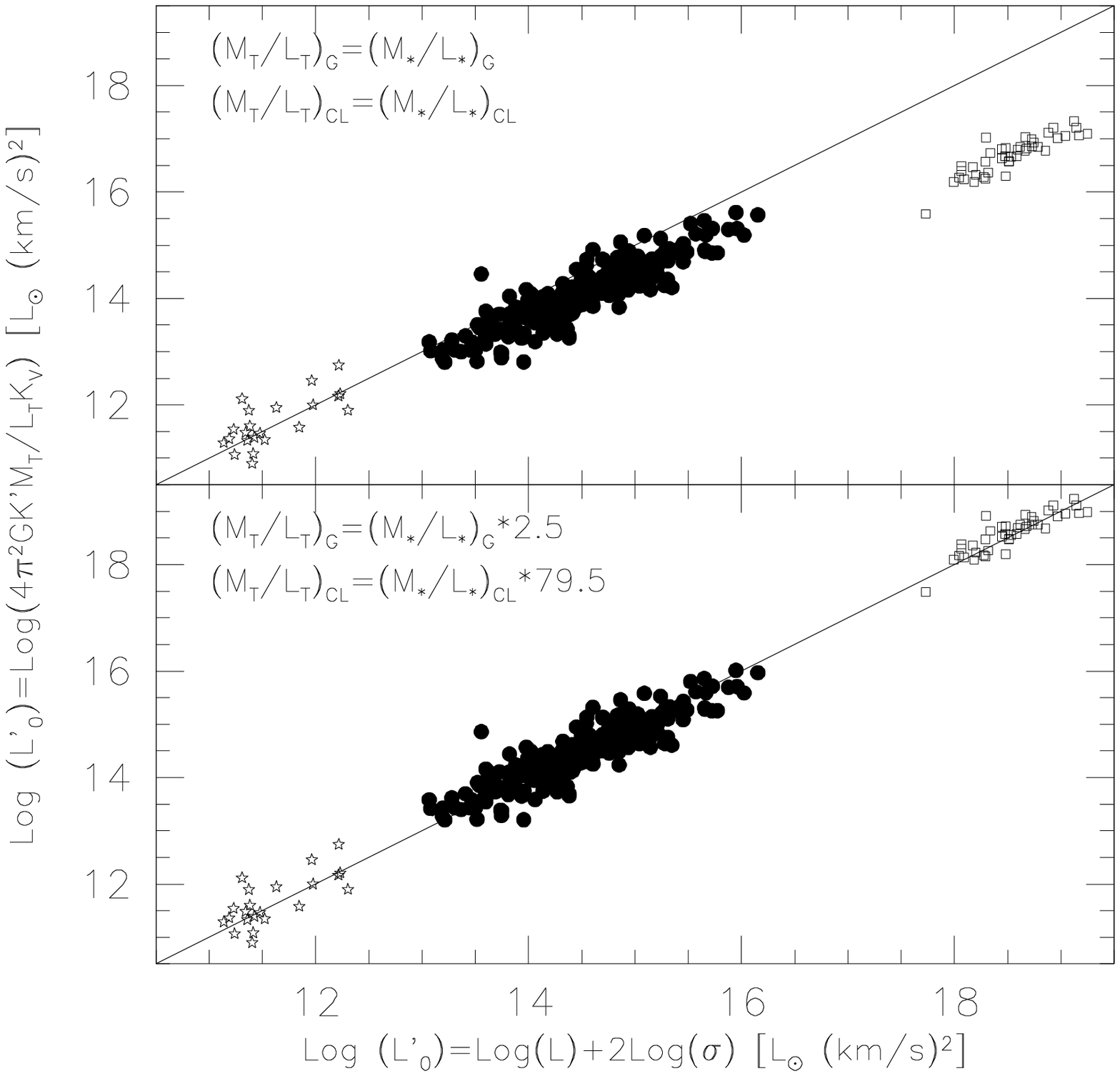}
\caption{Plot of the quantity $L'_0$ derived from Eq. \ref{loki} and from the relation $L=L'_0\sigma^{-2}$.\label{Fig9}}
\end{figure}

We want also to note that the quantity $L'_0$ gives the opportunity of quantify the DM content of ETGs. Fig. \ref{Fig9} shows
the comparison of the values of $\log(L'_0)$ derived from Eq. \ref{loki} and from the relation $L=L'_0\sigma^{-2}$.
Since along the y-axis we have only the observed stellar $M^*/L$ ratio while along the x-axis a quantity
depending on the total galaxy mass, it is possible to see that going toward more massive systems a progressively
larger fraction of DM is required to get the equivalence between the two quantities\footnote{Here the gas contribution to the 
total mass of the system is not taken into account}.

The nature of the $L_0$ and $L'_0$ parameters remains however quite elusive. Behind them is encripted the complex interconnection between the
SF process and the galaxy dynamics.

We note that Eqs. (\ref{loki}) and (\ref{lostar}) give consistently for $L_0$ a mean value of $1.6\pm2.8 \times 10^{29}$ $[gr/sec]$.
Where this value come from? We explored if the observed value of $L_0$ could be linked to the amount of matter burned by nuclear 
reactions inside stars in each second. Considering in fact only the H-burning of main sequence stars, we know that $\sim10^{18\div19}$ $[erg]$ 
comes from the nuclear
burning of 1 $gr$ of Hydrogen, so that at least $\sim10^{26\div27}$ $[gr]$ are burned each second in a galaxy with $10^{12}$ 
stars similarto the Sun. The remaining contribution can be easily explained taking into account that the light from a galaxy comes 
prevalently from RGB stars $\sim1000$ more luminous than the Sun which are burning Hydrogen in shells. This however rises the problem 
of explaining why $L_0$ is the same for galaxies of different masses (from $10^9$ to $10^{12}$ solar masses). Eq. (\ref{SFR2}) offers a more credible explanation 
for this: big/small systems have a larger/lower mean SFR and also a larger/lower velocity dispersion in such a way that the two compensate 
each other providing similar values of $L_0$ for many galaxies. We will see in Appendix how it is possible to achieve the observed values
of $L_0$ by looking at the FJ relation alon a different perspective.

$L'_0$ on the other hand is peculiar for each galaxy
and in some way can be considered a proxy of the so-called downsizing phenomenon, remembering the very different SFH of each galaxy.

In summary we can say that we have two different independent correlations. The first one is that between mass $M$ and velocity dispersion $\sigma$
provided by the virial theorem. The second one is that between luminosity $L$ and mean SFR $<SFR>$. Once we substitute $M$ with $L$ in the virial relation
we get the FJ relation, where $L_0=R_eL/GM$ (that is equal to $\sim 1.6\times10^{29}$ $gr/s$). With such substitution when we look at the 3D space
provided by $L-\sigma-<SFR>$ (in log units) we observe a sigar-shape distribution and we create a link between $\sigma$ and $<SFR>$ that are indirectly correlated.

With this in mind we now understand why we should use the $L=L'_0\sigma^{-2}$ relation for building the second virtual plane in the \muerespace\ space.
In fact in order to build such plane we need to use the direct correlation between $L$ and $<SFR>$ valid for each galaxy and not the one between $L$ and $\sigma$ valid
for all galaxies. This because we want to express the galaxy luminosity in a way independent on its mass.
The $L-<SFR>$ relation has $\sigma$ as second hidden parameter as we have seen.

In the next section we will further explore the consequences of
our findings for the problem of the star formation activity in galaxies.

\section{The SF activity in galaxies}\label{sec5}

Eq. (\ref{SFR2}) provides a link between $L_0$ and the mean SFR of galaxies. It does not give a direct link between the current SFR,
the velocity dispersion and $L_0$.

What we are looking for is a more direct link between these quantities. How are they connected?
We will show in Appendix that the FJ relation can be interpreted
as a possible translation of the Stefan-Boltzmann' law valid for stars to the case of stellar system, putting in evidence that it is 
always possible to express the energy of a system with the more convenient units (the ones we can measure).

Doing this exercise we have noted that
the galaxy luminosity can always be rewritten as:
\begin{equation}
L_G=  <\alpha_s>    N_s  < M_s v_s^{2} >,
\end{equation}
where the quantities within $<>$ are weighted averages over the whole stellar population. Here $N_s$ is the number of stars in the galaxy, $M_s$ 
their mass and $v^2_s$ their velocity dispersion. The constant $\alpha_s$ is different for each galaxy and represents the ratio between the total
energy emitted
in the form of electromagnetic radiation and the total kinetic energy of the galaxy.

In this context the quantity $L_0$ can be expressed by the relation:
\begin{equation}
 \frac{L_0}{\alpha_s} = M_g = \int_{0}^{t} \Psi(t) dt.
 \label{eqL0}
\end{equation}
where we have explicitly written the mass of the galaxy as the integral of its star formation rate and we have highlighted the dependence on time
of this parameter.

We can now recast Eqs. (\ref{loki}) and (\ref{eqL0}) in a different way putting in evidence the star formation rate of a galaxy. From this expression we can argue 
that at any epoch $t$ after virialization the SFR could be given by:

\begin{equation}
\Psi(t)=\frac{d}{dt}\left(\frac{4\pi^2 GK'(t)}{\alpha_s(t)K_V(t)}\frac{M(t)}{L(t)}\frac{1}{\sigma^4(t)}\right).
\label{sfreq}
\end{equation}

This relation is an important result because it allows the derivation of the SFR taking into account the projections of the FP
in the \muerespace\ space in a way consistent with observations. In particular we reproduce the ZOE limit in the observed distribution.
If we were using the original FJ Eq. (\ref{LFJ}) with $\beta=2$ we would obtain a SFR as a function of time that is proportional to the inverse square
of $\sigma$. This however would not be consistent with the observed distribution of galaxies in the projections of the FP.

Furthermore the relation of Eq. (\ref{sfreq}) tells us that at each cosmic epoch the SFR in a galaxy is not free. A galaxy can form  stars only at the rate
permitted by Eq. (\ref{sfreq}) along the whole cosmic history after virialization. In other words once the mass and the potential well of a galaxy is given,  
the star formation can go on according to the galaxy dynamics and to stellar evolution.

If a galaxy does not merge with others and does not experience a significant infall of new gas, its SFR will not be modified considerably continuing its evolution
according to Eq. (\ref{sfreq}).
Notably when the velocity dispersion is high the SFR is low. This is observed in today ETGs dominated by an high value of
$\sigma$ and an almost null SFR. Eq. (\ref{sfreq}) also tells us that when the parameters entering the relation do not change significantly (late stage of evolution)
the SFR is naturally around zero, providing a natural explanation for the mass-quenching phenomenon.

\begin{figure}
\plotone{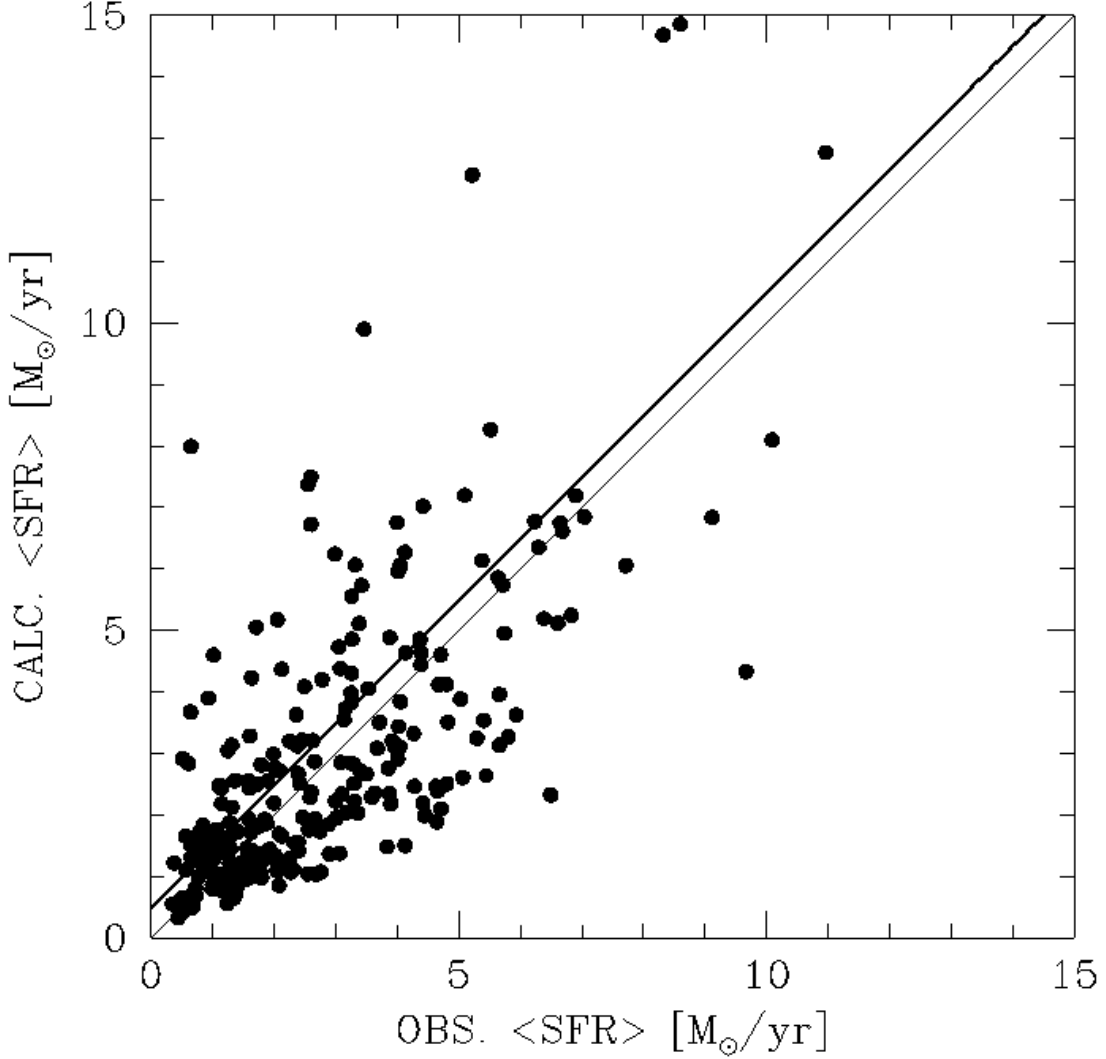}
\caption{Plot of the mean observed SFR measured by Fritz et al. (2007) for the galaxies of the WINGS database using the fitted spectral energy 
distributions versus the mean $<SFR>$ calculated on the basis of the prediction of Eq. (\ref{sfreq}) (see text). The thin line is the one-to-one 
relation, while the thick line is the fitted distribution.\label{Fig10}}
\end{figure}

Could we test in some way our prediction through observations?
Unfortunately testing the validity of Eq. (\ref{sfreq}) would require a database of masses, luminosities and velocity dispersions at different 
redshifts, while our WINGS database is made by nearby ETGs only. Considering that at $t=0$ the SFR was 0, we can only predict that the mean SFR of
today galaxies will be approximately given by:

\begin{equation}
<\Psi>=\frac{L_0}{\alpha_s\Delta t}\sim\frac{1}{2}\frac{M_G}{T_G},
\label{meansfr}
\end{equation}

where $\Delta t=T_G$ is the luminosity weighted age of the galaxies.

Fig. \ref{Fig10} shows the mean SFR measured by \cite{Fritz1}
in 4 distinct epochs from the direct fit of the galaxy SEDs
versus the mean SFR obtained by Eq. (\ref{meansfr}). The correlation ($c.c.=0.6$ and $rms\sim2.8$ but significant at a $\sim7\sigma$ confidence level) appears consistent with the theoretical expectation, taking into account the various 
sources of errors affecting both quantities, even if the sample is biased and the correlation may be driven by
few points at high SFR values. 

Eq. (\ref{sfreq}) is probably one key to understand the FP and the star formation history in galaxies. Its actual form depends on the original choice
made to formulate the FP problem, but probably behind such relation there is a sort of balance
equilibrium among the different forms of energies that are involved in the stellar systems (virial energy, radiation, non baryonic energy,
magnetic fields,..). For this reason the next step forward to understand these problems
could only come from detailed simulations of galaxy formation and evolution.

\section{Conclusions}\label{sec6}

We have shown that the origin of the FP can be traced back to the validity of two
basic physical relations: the virial dynamical equilibrium and the $L=L'_0\sigma^{-2}$ relation, which is a relation able to explain the existence of 
the downsizing phenomenon and the nature of the ZOE in the $I_e-R_e$ plane.

The galaxy luminosity could be correlated in two different ways with the SFR and the velocity dispersion, \ie\ through the $L=L'_0\sigma^{-2}$ relation 
and the $L=L_0\sigma^{2}$ relation. The first one is valid for a single galaxy, in the sense that $L'_0$ is very different for each object, while the second relation is valid for objects of quite different masses
(approximately from $10^9$ to $10^{12}$ $M_{\odot}$). 
Both relations have inside them a link with the SFR of galaxies. In the first one the primary role is that played by the SFR/SFH while the velocity dispersion enters
as a second less important parameter affecting in some way the SF. On the other hand when the primary role is played by the velocity dispersion we observe
that the residuals depend on the SFR (through the $M/L$ ratio, the effective surface brightness, etc.).

Since, as demonstrated by \cite{Zaritsky}, a Fundamental Manifold can be constructed for all stellar systems, the easy prediction is that in 
general the FP and FJ relations are different for each class of stellar system (GCs, dwarf galaxies, late-type galaxies, normal ETGs, cluster 
of galaxies). The diversity
is originated by the different zero-points of the VP and $L=L'_0\sigma^{-2}$ planes, or in other words by the different SFH and the different 
coupling between structure, dynamics and stellar populations.

The combination of the virial equilibrium, of the $L=L'_0\sigma^{-2}$ relation and the validity of the PFJ law for galaxy systems constrain 
objects of similar characteristics to the same FP, which is the locus of constant
$M/L$ ratio, $K_V$ and $SFR$ at each time epoch.

The projection of the intersecting lines connecting the VP and $L=L'_0\sigma^{-2}$ planes explains the properties observed for ETGs in the \muere\ plane,
in particular the existence of the ZOE that in this framework is the natural limit reached by the stellar and dynamical evolution
of a stellar system today.

The ZP of the FP provides a natural constraint to the possible SFR activity of a galaxy that at any given epoch
could not deviate from the track imposed by the mass and luminosity evolution in a given potential well. Conversely,
this relation provides the fine-tuning required to keep galaxies on a tilted FP with a small scatter.

Eq. (\ref{sfreq}) should be studied now through photometric and dynamical simulations following the
details of the mass assembly in stars and their relative luminosities.
Naively, we can predict
that since the stellar mass is generally increasing, while luminosity and stellar velocity dispersions could vary with the generations of stars, 
the resulting SFR will probably see various peaks at different redshift epochs depending on the galaxy dynamics.

It will be interesting to see if Eq. (\ref{sfreq}) will help to quantify the problem of the
star formation across the cosmic epochs and constrain in some way the mass quenching phenomenon.
First, it will be important to verify if the two principal
types of galaxies in the color - magnitude
(or stellar mass), color - concentration, and color - morphology diagrams
\citep{Stratevaetal2001,Kauffmannetal2003,Brinchmannetal2004,Baldryetal2004,Baldryetal2006,Driveretal2006,
Bamfordetal2009} can be reproduced. We know that in these plots there are two main regions: the so-called
‘blue cloud’ (or main sequence), where galaxy mass correlates
with the star formation rate, and the ‘red sequence’
where there is no such correlation and galaxies are passive.
The origin of this bi-modality is commonly attributed to the
bulge and disk structure of galaxies.
In general disks are bluer in color than bulges \citep[e.g.,][]{PeletierBalcells1996} and galaxies with
lower stellar mass and lower Sersic index
tend to be bluer (and hence have higher sSFRs) than higher stellar
mass and higher Sersic index systems \citep{Baldryetal2004,Driveretal2006,Baldryetal2006,Bamfordetal2009}.
Similar trends
are observed for luminosity and stellar light concentration
\citep{Stratevaetal2001,Driveretal2006}.
This idea fits with the found dependence of the SFR on the Sersic index and velocity dispersion found here.
Unfortunately all such relationships
are complicated by the effects of the environment, so that disentangling the various effects on the star
formation efficiency is quite difficult.

\appendix
\section{A possible origin for the FJ relation}
We try to demonstrate here that
the FJ relation could be seen as a sort of translation of the Black Body Stefan-Boltzmann 
law valid for individual stars to the case of a galaxy made by an assembly of stars in 
which the temperature is replaced by the velocity dispersion. From this analysis it will 
appear again the link connecting $L_0$ and $L'_0$ with the SFR of galaxies.

It goes without saying that there is not an immediate straight correlation between the 
physical situations in stars and galaxies; however, we will convincingly see that such 
analogy is possible and also argue that dynamics (via the velocity dispersion) and 
stellar populations in a galaxy (via the light emitted by stars) are each other intimately 
related. To demonstrate that this is possible we proceed as follow.

\subsection{The case of single stars}

A star of mass $M_s$, radius $R_s$, luminosity $L_s$, and effective temperature 
$T_{s,e}$ is an assembly of N heavy particles (nuclei, ions, and atoms, whereas
 electrons can be neglected) 
in thermal motion with mean temperature  $<T>$ and in virial equilibrium, i.e. satisfying the 
condition:

 \begin{equation}
   M_{s} v_{s}^2  \equiv \left| \frac {GM_{s}^2} {R_{s}}\right| \equiv E_V
 \end{equation}\label{virial}
where $v_s$ is the mean particle velocity in a gram
of matter,  $M_s= N<m_p>$ with N is the number of heavy particles and $m_p$ their  mean mass, 
and finally  $E_V$  stands for the "virial energy".

Consider first
the total bolometric luminosity of a star (i.e. the total energy emitted per second 
by the surface). This is usually derived from the Stefan-Boltzmann law, 
since stars are in good approximation Black Body systems.

In a star we can measure the luminosity $L_s$, the effective temperature $T_{es}$, and 
the radius $R_s$ which are related by the well known Black Body law 
($L = 4\pi R^2 \sigma_{SB} T^4$), where $\sigma_{SB}$ is the Stefan-Boltzmann constant. 
The suffix $SB$ is to distinguish it from the velocity dispersion of stars in a galaxy, 
usually indicated with the same symbol. When misunderstanding is obviously avoided, 
the suffix is dropped.

It is worth recalling here that the luminosity can be derived from the energy 
content of the Black Body  according to:

\begin{equation}
 U_{bb}(T)=\frac{8\pi \Omega}{h^3 c^3}(kT)^4\frac{\pi^4}{15}
\end{equation}\label{U_T}
where $\Omega$ is the total volume and $U_{bb}(T)$ the total energy of the Black Body. 
From this we obtain the luminosity of the  star
\begin{equation}
 L_s=\frac{U_{bb}(T)}{\Omega}4\pi R_s^2 c = \frac{3 U_{bb}(T)  c} {R_s}.
\end{equation}\label{L_S}

At this point we verify  that the gravitational energy, the  mean kinetic energy 
of the particles, and the Black Body energy content of the whole star with mean 
temperature $<T>$  are comparable to each other. Taking the Sun as a typical 
star, for which we assume $R_s = 6.94 \, 10^{10}$  cm, $M_s = 1.99 \, 10^{33}$ g, mean 
internal temperature $<T> \simeq 5 \times 10^6$ K, and central value   $T \simeq 10^7$,    
\footnote{The elementary theory of stellar evolution by combining the equations for 
 hydrostatic equilibrium, mass conservation and physical state of the plasma, e.g.
 $P = \frac{k}{\mu m_H}\rho T$,  provides a simple relation for the mean temperature 
inside a star
$$  \bar{T} \geq  4.58 \times 10^6  \mu \frac{M}{M_\odot} \frac{R_\odot}{R}\, \rm K$$
where $M$ and $R$ are the total mass and radius of the star and $\mu$ the mean molecular 
weight of the gas. For a solar like star $\mu \simeq 1$, so that $\bar T \simeq 5\times 10^6$.
The central temperature is higher than this and close to $10^7$. }
we obtain:

i) the mean density of kinetic energy  of the $N$ particles in the star is 

\begin{equation}
< E_k  >= \frac{1}{\Omega}\sum^N_i \frac{m_p v_p^2}{2}= \frac{3}{2}nKT \simeq 3.47
\times 10^{15}  \, \rm erg/cm^3
\end{equation}\label{E_K}

\noindent
where $m_p$ and $v_p$ are the mass and velocity of each particle and $n=N/\Omega$ is the 
number density of particles.

ii) the mean density of gravitational energy is
\begin{equation}
 <E_g>=\frac{ 3 G M_s^2}{ 4\pi R_s^4}  \simeq 2.72 \times 10^{15} \, \rm erg/cm^3
\end{equation}\label{E_G}

iii)  the  mean energy density  of  the photons is
\begin{equation}
 <U_{bb}(T)>=\frac{8 \pi^5}{15 h^3 c^3}(KT)^4 \simeq  1.15 \times 10^{15}\, \rm  erg/cm^3
\end{equation}\label{Mean_U}
 for a mean temperature of $10^7$ K. Within the 
numerical approximation the three energies are  of the same order. Strictly speaking 
one should have  $<E_g> \simeq <E_k> + <U_{bb}>$. Within the approximation our estimates 
fulfill this constraint. Analogous estimates 
can be made for other types of star with similar conclusions.  In other words , there seems to be 
a relationship between the gravitational energy density and the sum of the
electromagnetic and kinetic energy densities. 
Finally,
using the virial condition we can also estimate  the mean velocities of the 
particles in a star (the Sun in this example)  which are about 
$v_s \simeq 200$ km/s, depending on the exact value adopted for the temperature.

Given these premises, the luminosity of a star can be derived from

\begin{equation}
L_s= \left| \frac{dE_i}{dt}\right| \, ,
\end{equation}\label{lum_u}
where  $E_i$ is the total internal energy (sum of the nuclear and gravo-thermal contributions).
We may generalize the above relation by supposing that the luminosity can be expressed as:
\begin{equation}
 L_s = \alpha_s E_V  \equiv \alpha_s M_s <v_p>^2 \equiv \alpha_s  <U_{bb}>  
\frac{4}{3} \pi R_s^3
 \end{equation}\label{L_equiv}
where $\alpha_s$ is a suitable proportionality factor with the dimension of an inverse 
of time. In other words we link the luminosity $L_s$ to the internal properties of 
the star, in particular to the mean velocity of the constituent heavy particles.

However the same luminosity can be expressed by means of the surface Black Body  with 
temperature equal to the effective temperature $T_e$ of the star (a few thousands degrees, 
about $5.78\times10^3$ K for the Sun and $3 \times 10^3$ for a RGB star).

 \begin{equation}
 L= <U'_{bb}> 4 \pi c R_s^2
\end{equation}\label{U_primo}
where $U'_{bb}$ is the energy  of the black body at the surface  temperature. This implies that
the ratio of the external to the internal Black Body energies  is $U'_{bb} \simeq 10^{-13} U_{bb}$.  
The size of the proportionality coefficient can be understood  as due to the $T^4$ 
dependence of the Black Body energy density and the natural variation of the temperature 
from the surface to the inner regions of a star. The typical temperature gradient of a 
Sun like star is $\| \Delta T/ \Delta R\| \simeq 10^{-4} \, \rm K\, cm^{-1}$, 
 where $\Delta T = T-T' \simeq T$ and $\| \Delta R \| = \| R -R' \| \simeq R'$ if R and 
T refer to a inner region (close to the center) and $R'$ and $T'$ to the surface. 
Therefore  $T'/T \simeq 10^{-4}$.
\footnote{The values assumed for the central and surface temperature of the Sun amply justify 
a  ratio $T' /T \simeq 0.0001$  or lower and  a   proportionality factor 
$10^{-13}$ in the relationship between the energy densities $U_{bb}$ and $U'_{bb}$. }

It follows from all this that $\alpha_s\simeq 10^{-14}c/R_s$. Inserting 
the value for the light velocity and the radius of a typical star (like the Sun) 
one obtains $\alpha_s  \simeq 10^{-14} \, s^{-1}$.

 The factor $c/R_s$ secures that the energy density is translated to energy lost 
per unit time (a power). What we have done so far is a simple rephrasing of the 
classical expression for the luminosity. The reason for writing the star luminosity 
in this curious way will appear clear as soon as we move to galaxies, \ie\ to systems 
hosting billion of stars.

\begin{figure}
\plotone{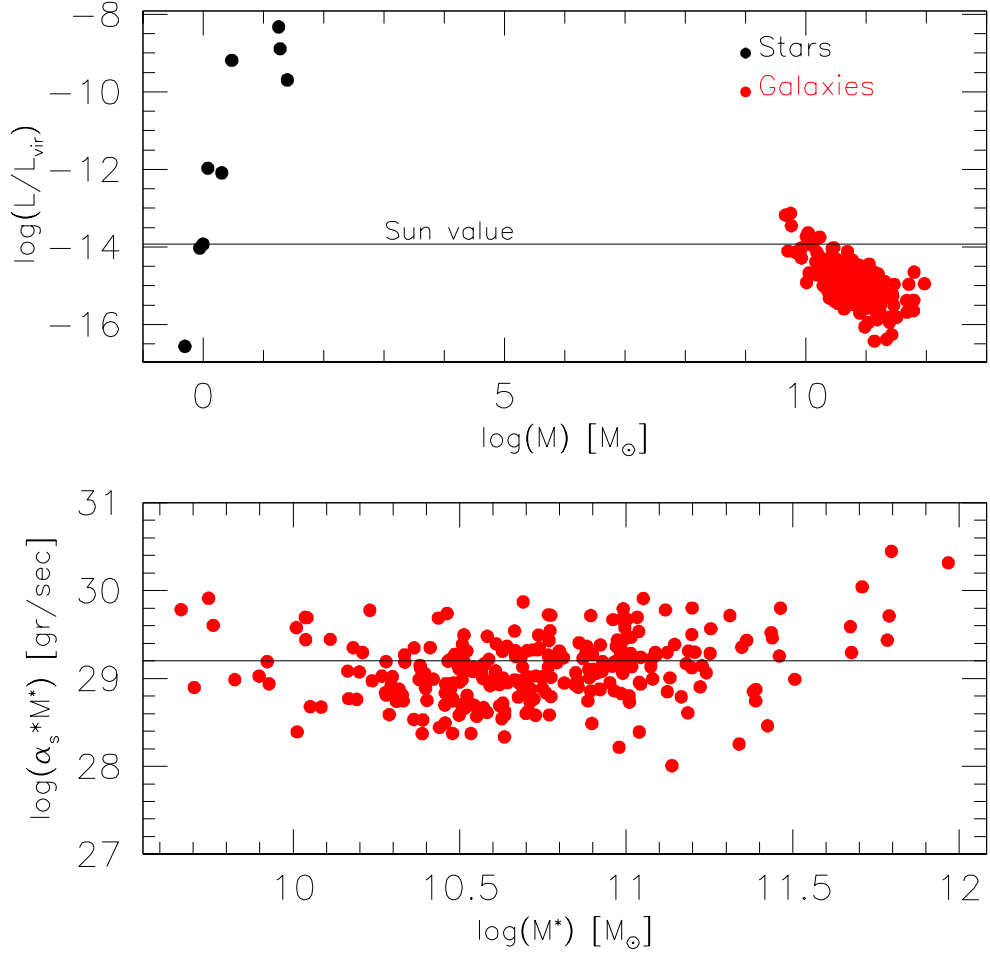}
\caption{Upper panel: Plot of  $\alpha_s$ for the stars and its analog for the galaxies $\alpha_G$ 
(see the text for the definition) as a function  the mass of the 
virialized system (star or galaxy as appropriate).  Note the large range of values 
spanned by $\alpha_s$ at varying the
mass of the star from a dwarf to a massive object. Finally note that the values $\alpha_G$ 
for the galaxies fall in range typical of  the low mass (old) stars. Lower panel:
plot of $L_0=\alpha_s*M_G$ (using $M^*$ instead of $M_G$) versus the galaxy masses.
The solid line gives the value observed for $L_0$ in the FJ relation.\label{FigA1}}
\end{figure}

The whole discussion above has been checked against stars like the Sun, so that one 
expect that changing
type of stars the value of $\alpha_s$ should change. This is shown in Fig. \ref{FigA1}. 
As expected $\alpha_s$ spans a wide range 
passing from dwarfs to massive stars,  but this will not 
affect  our final conclusions.

\subsection{The case of galaxies}

We extend now the above consideration and formalism to the case of a galaxy with 
mass $M_G$ and radius $R_G$, a large assembly of stars each of which shining with 
the luminosity $L_{s,i}$.
In brief, the  luminosity of  the galaxy  is the sum of the luminosity of the all 
the stars inside; the luminosity of each star can be expressed as proportional to 
the total kinetic energy of gas particles.  Therefore we may write

\begin{equation}
L_{G}=\sum^{N_s}_{i=1}\alpha_{s,i} M_{s,i} v_{s,i}^{2},
\end{equation}
where $N_s$ is the total number of stars within the galaxy, and $\alpha_{s,i}$, 
$M_{s,i}$,  $v_{s,i}$, and $R_{s,i}$ are the basic quantities characterizing each star.
In analogy with eq. (\ref{L_equiv}), the galaxy luminosity can be rewritten as:

\begin{equation}
L_G=  <\alpha_s>    N_s  < M_s v_s^{2}   >,
\end{equation}
where the quantities within $<>$ are weighted averages over the whole stellar population.
Note that for galaxies of the same "size" (mass and radius) these values will be 
very similar. 

Now,  thanks to the homologous nature of the gravitational collapse at all scales, it 
is possible to note that
the quantity $<v_s>$, i.e. the mean velocity of particles inside a star, turns out to 
be comparable to the velocity dispersion of stars within a galaxy, customarily named  
 $\sigma$ (in km/sec).
It is then possible to write:

\begin{equation}
L=L_0\sigma^2,
\end{equation}
where

\begin{equation}\label{lostar}
L_0=<\alpha_s>   N_s <M_s> \equiv <\alpha_s> M_G.
\end{equation}

\noindent
It can be shown that even for a galaxy there is a 
relationship (by chance?) between the 
total gravitational energy, the total kinetic energy of the stars,
  and total radiative energy emitted by stars so that the 
relation (\ref{lostar}) can be replaced by

\begin{equation}\label{lostar1}
L_0=<\alpha_{G}> M_{G} \equiv \frac{c}{R_{G}} M_{G}
\end{equation}
where $\alpha_{G}$ refers to the  galaxy as a whole. 
Like in the case of stars, $<\alpha_G>$ has the dimension of 
an inverse of time.

To demonstrate the validity of Eq. (\ref{lostar1}) we consider a generic 
mean stellar content of
$N_s \simeq 10^{12}$ objects for simplicity taken like to the Sun 
($M_\odot = 2\times 10^{33} $ g and 
radius $R_\odot = 6.94 \times 10^{10}$ cm, surface temperature $T_s \simeq 5780$ K), total
 mass $M_G = 10^{12}\, M_\odot$, 
total radius $R_G \simeq 100$ kpc. In this example we ignore the contribution to the mass 
given by Dark Matter (DM). According to the current understanding of the presence of DM 
in galaxies, the ratios of the dark to baryonic matter (BM)  both in mass and radii of the 
spatial distributions (supposed to spherical)  $M_{DM} \simeq \beta \times M_{BM}$ and 
$R_{DM}= \beta \times R_{BM}$. This means that within the volume occupied by the BM  there is 
about  $1/\beta^2 \times M_{DM}$ \citep{Bertinetal1992,Sagliaetal1992,Bertin}. 
For current estimates of $\beta \simeq 6$ DM 
can be neglected  in the internal regions of a galaxy where stars are located.

\begin{figure}
\plottwo{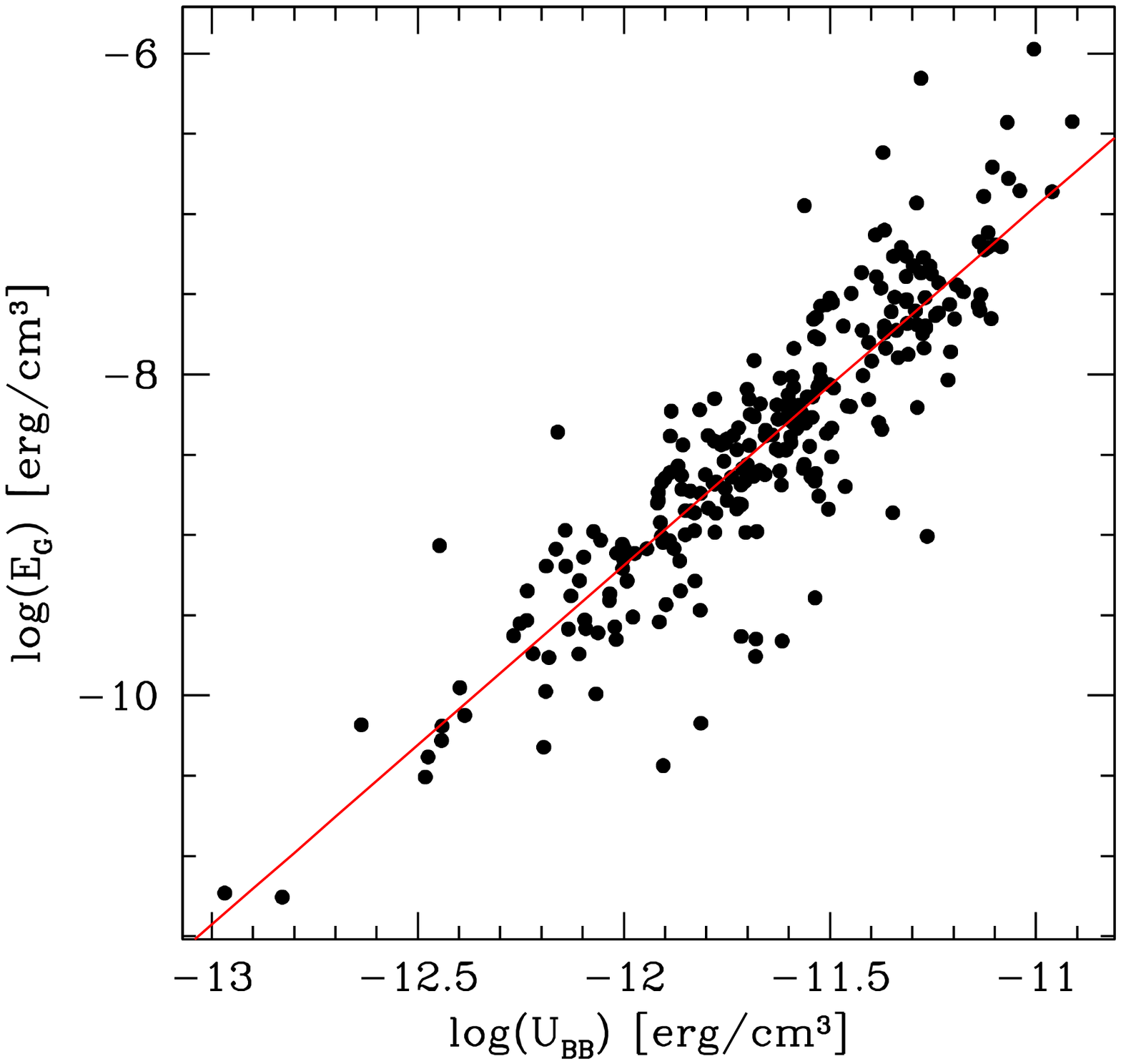}{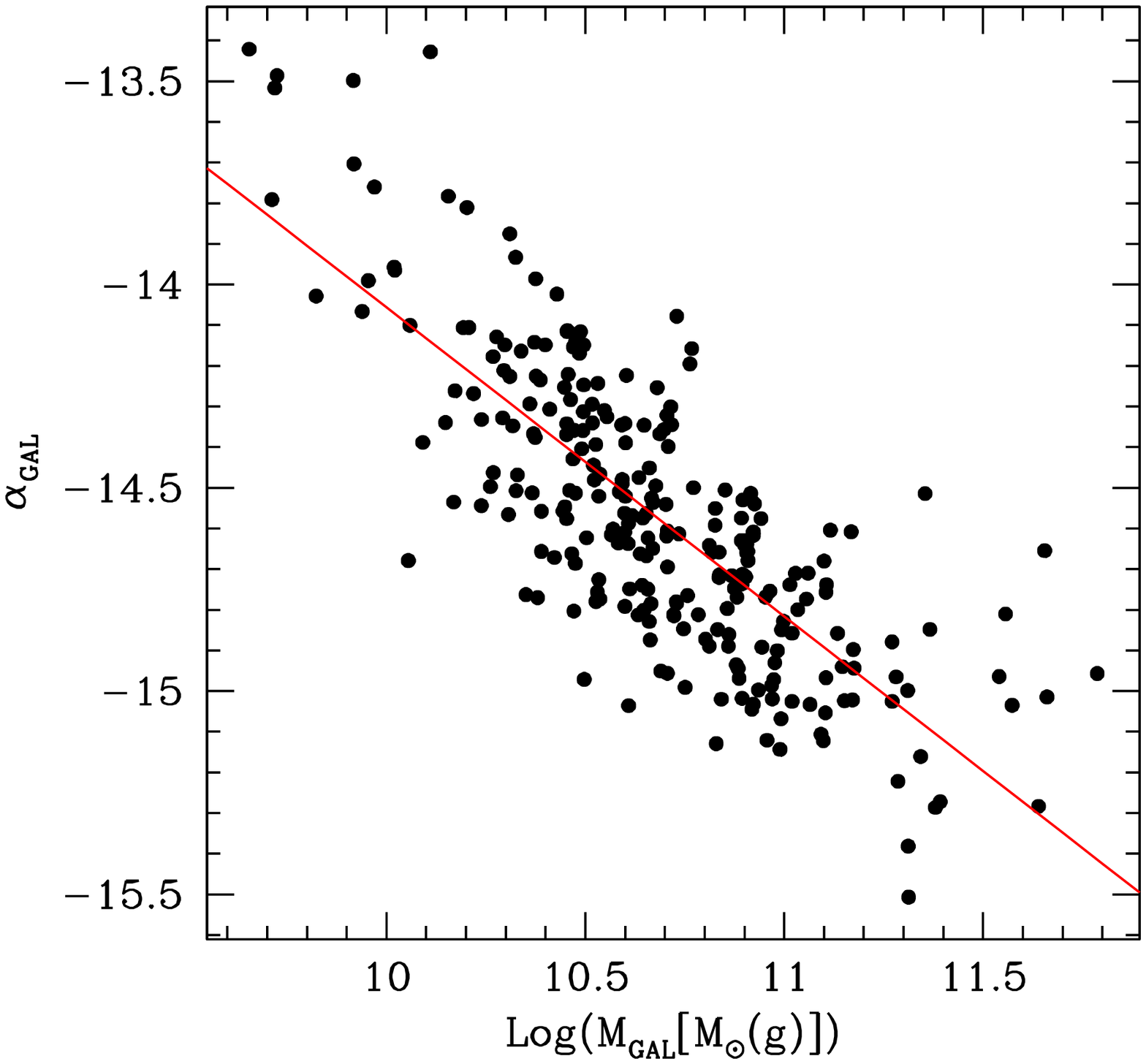}
\caption{Left panel: The mean density of the gravitational energy versus the mean density of 
the BB energy for a sample of early type galaxies. 
Right Panel: the quantity $\alpha_G$ as function of  the stellar galaxy mass in solar units 
for the object of the same sample.\label{FigA2}}
\end{figure}

The energy density
of  the photons emitted by all the stars in the galaxy  evaluated at any arbitrary point 
inside the galaxy is given by
\begin{equation}\label{ugal}
 U_{bb,G}=\int_0^{R_{G}}  U'_{bb,s}  \frac{N_s}{\Omega_{G}} \quad 4 \pi r^2 dr \frac{R_s^2}{r^2}
\end{equation}
where   $U'_{bb,s}$ refers to the Black Body at the temperature of the stellar sources
\footnote{In relation to this, we remind the reader that in most galaxies nearby the detected
 light is due to stars from the main sequence turnoff (or slightly fainter than) to the tip 
of the RGB. In sufficiently old galaxies the corresponding mass range is rather small. 
In other words, the stellar population responsible for the observed light can be reduced 
to a single population of a certain age and mean chemical composition.  },
$\Omega_G$ is the volume of the whole galaxy, and the factor $U'_{bb,s} \times N_s / \Omega_G$ 
the mean Black Body radiation inside the galaxy.
Although the integrand of Eq. (\ref{ugal}) is not strictly correct to evaluate the
 variation of the Black Body energy as a function of the galacto-centric distance, it is 
adequate to our purposes. The quantity $U'_{bb,s}$ is given by

\begin{equation}
U'_{bb,s}=\frac{8\pi}{h^3 c^3}\frac{\pi^4}{15}(k T)^4  \simeq 8.02   \,\, \rm erg \, cm^{-3}
\end{equation}
\noindent
so that for $U_{bb,G}$ of eq. (\ref{ugal}) we estimate
\begin{equation}
 U_{bb,G}\simeq   1.29 \times 10^{-12}   \,\, \rm erg \, cm^{-3}
\end{equation}
\noindent
where we assumed $N_s \simeq 10^{12}$ stars,  $R_s \simeq 6.94 \times 10^{10}$ cm 
(roughly the solar radius), $\Omega_G \simeq 1.13 \times 10^{71}$ cm$^3$ for a 
galactic radius of about 100 kpc. 

The mean  density of kinetic energy of the stars  turns out to be of the order of 
$3.52 \times 10^{-12}\,\, \rm erg \,cm^{-3}$ for a mean velocity dispersion of about 
200 km $\rm s^{-1}$.

The mean gravitational energy density for the galaxy (limited to the volume occupied by the 
BM) is
\begin{equation}
 E_{g,G}=G \frac{M_{G}^2}{R_{G}^4}\frac{3}{4\pi}\simeq 
                   7.79 \times 10^{-12}   \,\, \rm erg \, cm^{-3}
\end{equation}
The gravitational energy is surely underestimated because we have neglected the presence of 
Dark Matter.

Therefore, also in this case there is an approximate relationship between the gravitational and 
the sum of electromagnetic 
and kinetic energy densities. 

We can then write the equation:
\begin{equation}
 L_{G}=\alpha_{G} M_{G} \sigma^2=\alpha_{G}<U_{G}> R_{G}^3 .
\end{equation}

Thanks to the assumption of uniform distribution of stars and stellar types most 
contributing to the light in  our model galaxy, also the  distribution of the photon
 energy inside  is uniform and always equal to the that of many black bodies of similar 
temperature. 
Furthermore, owing to the very large number of stars in a galaxy the light emitted by a 
certain region, e.g. within the effective radius, can be assimilated to that of black body 
of  certain mean temperature and very large surface. Therefore we may write
\begin{equation}
 L_{G}=<U_{bb,G}> 4 \pi c R_{G}^2
\end{equation}
so that for solar like stars 
$\alpha_{G}\simeq c/R_{G}\simeq 10^{-13} \, {\rm s ^{-1}} \simeq \alpha_s$.
It is worth emphasizing  here that $\alpha_G$ is nearly identical to $\alpha_s$.

In conclusion the classical Faber-Jackson relationship $L=L_0\sigma^2$ can be understood as
 a sort of translation of 
of Stefan-Boltzmann law for BBs to the case of
 galaxies that can be viewed as the sum of many BBs.

Fig. \ref{FigA1} shows the range  of values for the parameter $\alpha_G$  
of galaxies and compares them with those for stars. Note
that low mass galaxies have in general higher values of $\alpha$ (closer to the values  
for intermediate mass stars), whereas the big galaxies
are preferentially populated  by low mass stars.  What matters here  is that for every galaxy 
it exists a combination of $L_0$ ($\sim \alpha_s M_G$) and $\sigma$ able to 
reproduce the total galaxy luminosity. The lower panel of Fig. \ref{FigA1}
shows that $L_0=\alpha_s*M_G$ is approximately constant for wide range of galaxy masses. 

We have calculated $U_{bb,G}$ and $E_g$ for a small sample of early type galaxies 
\citep{Morettietal} for which 
all the basic data were available  and estimated the parameter 
$\alpha_G$ for all of them. The results are shown in the two panels of Fig. \ref{FigA2}.

One might argue whether this is true also for spiral galaxies.
We believe that the origin of the Tully-Fisher relation for late-type systems can be likely 
reported to the same context. Here the mean characteristic velocity of the stellar
 system is no longer the velocity dispersion, but the circular rotation. 
For more complex systems, where rotation and velocity dispersion are significant,
 a combination of the two is required to characterize the total kinetic energy. 
The issue, however, is left to a future investigation.




\begin{thebibliography}{}



\bibitem[Baldry et al.(2004)]{Baldryetal2004}Baldry I. K., Glazebrook K., Brinkmann J., Ivezic Z., Lupton R. H., Nichol R. C., Szalay A. S.\ 2004, \apj, 600, 681
\bibitem[Baldry et al.(2006)]{Baldryetal2006}Baldry I. K., Balogh M. L., Bower R. G., Glazebrook K., Nichol R. C., Bamford S. P., Budavari T.\ 2006, \mnras, 373, 469
\bibitem[Bamford et al.(2009)]{Bamfordetal2009}Bamford S. P. et al.\ 2009, \mnras, 393, 1324
\bibitem[Bender, Burstein \& Faber(1992)]{BBF}Bender R., Burstein D., Faber S.M.\ 1992, \apj, 399, 462
\bibitem[Bertin, Saglia, \& Stiavelli(1992)]{Bertinetal1992} Bertin G., Saglia R.~P., Stiavelli M.\ 1992, \apj, 384, 423 
\bibitem[Bertin et al.(2002)]{Bertin} Bertin G., Ciotti L., Del Principe M.\ 2002, \mnras, 386, 149
\bibitem[Borriello et al.(2001)]{Borriello}Borriello A., Salucci P., Danese L.\ 2001, \mnras, 341, 1109
\bibitem[Bolton et al.(2008)]{Bolton}Bolton A.S., Treu T,, Koopmans L.V. E., et al.\ 2008, \apj, 684, 248
\bibitem[Brinchmann et al.(2004)]{Brinchmannetal2004}Brinchmann J., Charlot S., White S. D. M., Tremonti C., Kauffmann G., et al.\ 2004, \mnras, 351, 1151
\bibitem[Burkert(1993)]{Burkert93}Burkert A.\ 1993, \aap, 278, 23
\bibitem[Burstein(1997)]{BBFN}Burstein D., Bender, R., Faber, S.M., Nolthenius, R.\ 1997, \aj, 114, 1365
\bibitem[Busarello et al.(1997)]{Busarello}Busarello G., Capaccioli M., Longo G., Puddu E.\ 1997, In: The Second Stromlo Symposium ``The nature of Elliptical Galaxies'', ASP Conference Series, 166, 184
\bibitem[Caon et al.(1993)]{Caon} Caon N., Capaccioli M., D'Onofrio M.\ 1993, \mnras, 265, 1013
\bibitem[Capaccioli(1987)]{Capaccioli87}Capaccioli M.\ 1987, In: Structure and dynamics of elliptical galaxies, ed. P.T. de Zeeuw (Reidel, Dordrecht), p. 47
\bibitem[Capaccioli(1989)]{Capaccioli89}Capaccioli M.\ 1989, In: The world of galaxies, ed. H.G. Corwin \& L. Bottinelli (Springer-Verlag, Berlin), p. 208
\bibitem[Cappellari et al.(2006)]{Cappellari}Cappellari M., Bacon R., Bureau M., et al.\ 2006, \mnras, 366, 1126
\bibitem[Cappellari et al.(2007)]{capp07}Cappellari M., Emsellem E., Bacon R., et al.\ 2007, \mnras, 379, 418
\bibitem[Cappellari et al.(2012)]{Cappellari2}Cappellari M., McDermid R.M., Alatalo K., et al.\ 2012, \nat, 484, 485
\bibitem[Cappellari et al.(2013)]{Cappellari3}Cappellari M., McDermid R.M., Alatalo K., et al.\ 2012, \mnras, 432, 1862
\bibitem[Chiosi et al.(1998)]{Chiosi}Chiosi C., Bressan A., Portinari L., Tantalo R.\ 1998, \aap, 339, 355
\bibitem[Chiosi \& Carraro(2002)]{ChiosiCarr}Chiosi C., Carraro G.\ 2002, \mnras, 335, 335
\bibitem[Ciotti, Lanzoni \& Renzini(1996)]{Ciotti}Ciotti L., Lanzoni B., Renzini A.\ 1996, \mnras, 282, 1
\bibitem[de Carvalho \& da Costa(1988)]{deCarvalho88}de Carvalho, R.R., da Costa L.N.\ 1988, \apjs, 68, 173
\bibitem[Dekel \& Cox(2006)]{DekelCox}Dekel A., Cox T.J.\ 2006, \mnras, 370, 1445
\bibitem[Desmond \& Wechsler(2016)]{Desmond}Desmond H., Wechsler R.H.\ 2016, 2016arXiv160404670D
\bibitem[Djorgovski \& Davis(1987)]{DjorgDavis}Djorgovski S., Davis M.\ 1987, \apj, 313, 59
\bibitem[Djorgovski et al.(1998)]{Djorgovski}Djorgovski S., De Carvalho R., Han S.M.\ 1988, ASPC, 4, 329
\bibitem[D'Onofrio et al.(2013)]{Donofrioetal}D'Onofrio M., Fasano G., Moretti A., Marziani P., et al.\ 2013, \mnras, 435, 45
\bibitem[Dressler et al.(1987)]{Dressetal}Dressler A., Lynden-Bell D., Burstein D., Davies R.L., Faber S.M., Terlevich R.J., Wegner G.\ 1987, \apj, 313, 42
\bibitem[Driver et al.(2006)]{Driveretal2006}Driver S. P. et al.\ 2006, \mnras, 368, 414
\bibitem[Faber et al.(1987)]{Faber87}Faber S.M., Dressler A., Davies R., Burstein D., Lynden-Bell D.\ 1987, In: Nearly normal galaxies: From the Planck time to the present; Proceedings of the Eighth Santa Cruz Summer Workshop in Astronomy and Astrophysics, Santa Cruz, CA, July 21-Aug. 1, 1986 (A88-18401 05-90). New York, Springer-Verlag, 1987, p. 175-183
\bibitem[Faber \& Jackson(1976)]{FaberJackson}Faber S.M., Jackson R.E.\ 1976, \apj, 204, 668
\bibitem[Forbes et al.(1998)]{Forbes}Forbes D.A., Ponman T.J., Brown R.J.N.\ 1998, \apj, 508, L43
\bibitem[Fritz et al.(2007)]{Fritz1} Fritz J., Poggianti B. M., Bettoni D., et al.\ 2007, \aap, 470, 137
\bibitem[Gargiulo et al.(2009)]{Gargiulo}Gargiulo A.\ 2009, \mnras, 397, 75
\bibitem[Gerhard et al.(2001)]{Gerhard}Gerhard O., Kronawitter A., Saglia R.P., Bender R.\ 2001, \aj, 121, 1936
\bibitem[Graham \& Colless(1997)]{GrahamColless}Graham A., Colless M.\ 1997, \mnras, 287, 221
\bibitem[Graves et al.(2009)]{Gravesetal}Graves G.J., Faber S.M., Schiavon R.P.\ 2009, \apj, 698, 1590
\bibitem[Hjorth \& Madsen(1995)]{Hjorth}Hjorth J., Madsen J.\ 1995, \apj, 445, 55
\bibitem[Hopkins et al.(2008)]{Hopkins}Hopkins Ph.F., Cox T.J., Hernquist L.\ 2008, \apj, 689, 17
\bibitem[Kauffmann et al.(2003)]{Kauffmannetal2003}Kauffmann G. et al.\ 2003, \mnras, 346, 1055
\bibitem[Kormendy(1977)]{Kormendy}Kormendy J.\ 1977, \apj, 218, 333
\bibitem[La Barbera et al.(2000)]{LaBarb}La Barbera F., Busarello G., Capaccioli M.\ 2000, \aap, 362, 851
\bibitem[La Barbera et al.(2010)]{LaBarbera}La Barbera F., de Carvalho R. R., de La Rosa I. G., Lopes P. A. A.\ 2010, \mnras, 408, 1335
\bibitem[Magoulas et al.(2012)]{Magoulas}Magoulas C., et al.\ 2012, \mnras, 427, 245
\bibitem[Michard(1985)]{Michard85}Michard R.\ 1985, \aaps, 59, 205
\bibitem[Moretti et al.(2014)]{Morettietal}Moretti A., et al.\ 2014, \aap, 564, 138
\bibitem[Nipoti et al.(2006)]{Nipoti}Nipoti C., Londrillo P, Ciotti L.\ 2006, \mnras, 370, 681
\bibitem[O{\~n}orbe et al.(2005)]{Onorbe}O{\~n}orbe J., Dom\'inguez-Tenreiro R., S\'aiz A., Serna A., Artal H.\ 2005, \apj, 632, L57
\bibitem[Pasquato \& Bertin(2008)]{Pasquato}Pasquato M., Bertin G.\ 2008, \aap, 489, 1079
\bibitem[Pahre et al.(1998)]{Pahre}Pahre M.A., De Carvalho R.R., Djorgovski S.G.\ 1998, \aj, 116, 1606
\bibitem[Peletier \& Balcells(1996)]{PeletierBalcells1996}Peletier R. F., Balcells M.\ 1996, \aj, 111, 2238
\bibitem[Prugniel \& Simien(1997)]{PrugSimien}Prugniel Ph., Simien F.\ 1997, \aap, 321, 111
\bibitem[Renzini \& Ciotti(1993)]{Renzini}Renzini A., Ciotti L.\ 1993, \apj, 416, L49
\bibitem[Robertson et al.(2006)]{Robertson}Robertson B., Cox T.J., Hernquist L., et al.\ 2006, \apj, 641, 21
\bibitem[Saglia, Bertin, \& Stiavelli(1992)]{Sagliaetal1992} Saglia R.~P., Bertin G., Stiavelli M., 1992, \apj, 384, 433 
\bibitem[Schombert(1986)]{Schombert86}Schombert J.M.\ 1986, \apjs, 60, 603
\bibitem[Scodeggio et al.(1998)]{Scodeggio}Scodeggio M., Gavazzi G., Belsole E., Pierini D., Boselli A.\ 1998, \mnras, 301, 1001
\bibitem[Secco(2001)]{Secco2001}Secco L.\ 2001, New Astr. 6, 339
\bibitem[Secco \& Bindoni(2009)]{Secco}Secco L, Bindoni D.\ 2009, New Astr. 14, 567
\bibitem[Strateva et al.(2001)]{Stratevaetal2001}Strateva I., et al.\ 2001, \aj, 122, 1861
\bibitem[Terlevich \& Forbes(2002)]{Terlevich}Terlevich A.I., Forbes D.A.\ 2002, \mnras, 330, 547
\bibitem[Tortora et al.(2009)]{Tortora}Tortora C., Napolitano N. R., Romanowsky A. J., Capaccioli M., Covone G.\ 2009, \mnras, 396, 1132
\bibitem[Treu et al.(2005)]{Treu05}Treu T., Ellis R.K., Liao T.X., van Dokkum P.G.\ 2005, \apj, 622, L5
\bibitem[Trujillo et al.(2004)]{Trujillo}Trujillo I., Burkert A., Bell E.F.\ 2004, \apj, 127, 1917
\bibitem[Tully \& Fisher(1977)]{TullyFisher}Tully B.R., Fisher J.R.\ 1977, \aap, 54, 661
\bibitem[Young \& Currie(1994)]{YoungCurrie} Young C.K. \& Currie M.J.\ 1994, \mnras, 268, 11
\bibitem[Zaritsky(2012)]{Zaritsky} Zaritsky D.\ 2012, ISRN Astronomy and Astrophysics, vol. 2012, id. 189625

\end{thebibliography}
\end{document}